\documentclass[prd,reprint,showpacs,superscriptaddress]{revtex4-1}
\usepackage{tikz,graphicx,booktabs,multirow,array,microtype,quark,mathtools,float}
\usepackage[colorlinks=true,backref=false, linktocpage=true,
citecolor=blue,
urlcolor=blue,linkcolor=blue,
pdfpagemode=UseOutlines]{hyperref}
\newcolumntype{C}{>{$}c<{$}}
\usepackage{dcolumn}
\AtBeginDocument{
\heavyrulewidth=.08em
\lightrulewidth=.05em
\cmidrulewidth=.03em
\belowrulesep=.65ex
\belowbottomsep=0pt
\aboverulesep=.4ex
\abovetopsep=0pt
\cmidrulesep=\doublerulesep
\cmidrulekern=.5em
\defaultaddspace=.5em
}
\def\fm {\mathop{\hbox{fm}}}
\def\MeV {\mathop{\hbox{MeV}}}
\def\GeV {\mathop{\hbox{GeV}}}

\def\Re {\mathop{\hbox{Re}}}

\def\Tr {\mathop{\hbox{Tr}}}

\def\erf {\mathop{\hbox{erf}}}

\usepackage{amsmath} 
\usepackage{dsfont}  
\usepackage{bm}      

\def\reals{\mathds{R}}
\def\beq{\begin{equation}}
\def\eeq{\end{equation}}
\def\beqs#1\eeqs{\beq\begin{split} #1 \end{split}\eeq}

\def\pmat#1{\begin{pmatrix}#1\end{pmatrix}}

\long\def\comment#1{}

\def\norm#1{\left\| #1 \right\|}
\def\bb{[\![}
\def\rr{]\!]}

\def\ket#1{\left| #1 \right\rangle}
\def\bra#1{\left\langle #1 \right|}

\def\opbraket#1#2#3{ \left\langle #1 \left| #2 \right| #3 \right\rangle}
\def\av#1{ \left\langle #1 \right\rangle }

\def\uchipt{U$\chi$PT}

\begin{document}
\title{Rho resonance parameters from lattice QCD}
\author{Dehua Guo}
\affiliation{Physics Department, The George Washington University, Washington, DC 20052, USA}
\email{guodehua@gwu.edu}
\author{Andrei Alexandru}
\email{aalexan@gwu.edu}
\affiliation{Physics Department, The George Washington University, Washington, DC 20052, USA}
\author{Raquel Molina}
\email{ramope71@gwu.edu}
\affiliation{Physics Department, The George Washington University, Washington, DC 20052, USA}
\author{Michael D\"oring}
\email{doring@gwu.edu}
\affiliation{Physics Department, The George Washington University, Washington, DC 20052, USA}
\affiliation{Thomas Jefferson National Accelerator Facility, Newport News, VA 23606, USA}
\date{\today}

\begin{abstract}
We perform a high-precision calculation of the phase shifts for $\pi$-$\pi$ scattering in the 
$I=1$, $J=1$ channel in the elastic region using elongated lattices with two mass-degenerate 
quark flavors ($N_f=2$). We extract the $\rho$ resonance parameters using a Breit-Wigner fit 
at two different quark masses, corresponding to $m_\pi=226\MeV$ and $m_\pi=315\MeV$, and perform 
an extrapolation to the physical point. The extrapolation is based on a unitarized chiral 
perturbation theory model that describes well the phase-shifts around the resonance for both quark masses.
We find that the extrapolated value, $m_\rho=720(1)(15)\MeV$, is significantly lower that 
the physical rho mass and we argue that this shift could be due to the absence of the strange quark in 
our calculation. 

\end{abstract}
\pacs{12.38.Gc}

\maketitle
\section{Introduction}


A large experimental and theoretical effort is dedicated to measuring scattering cross-sections in different channels, extracting phase-shifts, and determining the parameters for resonances. Lattice QCD calculations can be used to complement this efforts by providing input that is either difficult to measure directly or not accessible in experiments, for example by using non-physical quark masses which can help develop better phenomenological models. In this study, we will focus on the $\rho(770)$ resonance in the isospin-1, spin-1 channel for pion-pion scattering. This resonance was the subject of a number of lattice QCD calculations~\cite{Aoki:2011yj,Gockeler:2008kc,Feng:2010es,Lang:2011mn,Pelissier:2012pi, Bali:2015gji, Frison:2010ws, Aoki:2011yj, Dudek:2012xn, Metivet:2014bga, Wilson:2015dqa, Bulava:2016mks,Feng:2014gba}, with the results for phase-shifts becoming more and more precise and the quark masses getting closer to the physical point. 


Scattering information is determined from lattice QCD indirectly by computing the energy of the two hadron states in a finite box with periodic boundary conditions. In a series of papers L\"{u}scher derived a formula connecting the two-hadron energy
states with the phase shift, valid up to exponential corrections that decrease with volume~\cite{Luscher:1985dn,Luscher:1986pf,Luscher:1990ck,Luscher:1990ux}. The formula was originally developed for two particle states with total zero momentum in a cubic box and it was latter extended to non-zero momentum states (boosted frame)~\cite{Rummukainen:1995vs} and asymmetric (elongated) boxes~\cite{Feng:2004ua}. These extensions were developed to extract phase shifts in different kinematic regions, where the relative momentum of the scattering particles is smaller that in the original setup when using similar lattice volumes. In this study, we use elongated boxes and extract phase-shifts both for zero-momentum states and for boosted states, states with a moving center-of-mass, so that we can finely scan the phase shift pattern around the resonance region.

Our study is carried out using $N_f=2$ dynamical configurations with nHYP fermions~\cite{Hasenfratz:2007rf}. This study extends a previous calculation~\cite{Pelissier:2012pi} by adding a larger base of interpolator fields and a set of ensembles at lower pion masses. Using two different sea quark masses 
allows us to extrapolate to the physical point. The need for a larger interpolator field basis is discussed below.

To obtain the energy spectrum of a resonance in a specific scattering channel, the choice of the interpolating fields is nontrivial. The interpolating fields should not only have the correct symmetries but also have enough overlap with the relevant eigenstates of the system. For example, it is known (and our study confirms it) that the quark-antiquark ($q\bar{q}$) operators do not have enough overlap with the few lowest energy states in the channel studied in this paper and two-hadron interpolator fields are required. Moreover, as the pion mass becomes smaller, more multi-hadrons states appear near and below the resonance region. As a result, more and more multi-hadron operators have to be included in order to capture these states. For the channel studied in this paper, two-pion states with different scattering momentum have to be included to resolve the energy spectrum near the $\rho$ resonance region. 


The large interpolating field basis leads to a large number of correlation functions that need to be evaluated using lattice QCD techniques. The most computational demanding diagrams are the four-point correlation functions that arise from Wick contractions generated by the two-hadron interpolating fields. These diagrams require knowledge of the quark propagator from all points on the lattice to all other points. Direct evaluation of this {\em all-to-all propagator} is impractical. The standard techniques used to overcome this problem are stochastic evaluation or Laplacian-Heaviside (LapH) smearing~\cite{Peardon:2009gh}, which only require the evaluation of the quark propagator in a smaller subspace. In our previous study~\cite{Pelissier:2012pi} we used a stochastic method, but we decided to use LapH method in this study. The advantage of the latter method is that it separates the calculation of the quark propagator from the evaluation of the hadronic correlation functions and it allows more flexibility in constructing the interpolator basis. 


Note that in this study we use dynamical configurations with two mass-degenerated fermions ($N_f=2$) and the effects of the strange sea quarks are not included. This is not ideal if one wants to compare the results with the physical ones but it does have some advantages. First, it has the advantage that the L\"uscher equation can be used in the entire region below the $4m_\pi$ threshold because the $K\bar K$ channel is absent. Secondly, the results of this study allow us to gauge the impact of the strange quark on the properties of the $\rho$ resonance. As we will show, at least in some respects, the absence of the strange quark leads to surprisingly large effects.

We analyze two sets of ensembles with different sea quark masses: one corresponding to $m_{\pi}= 315\MeV$ and the other to $m_{\pi}=226\MeV$. For each pion mass we use three ensembles with different lattice geometry. For each ensemble we analyze states at rest $\mathbf{P}=(0,0,0)$, and states moving along the elongated direction with momentum $\mathbf{P}=(0,0,1)$. For each case, we use four different $q\bar{q}$ interpolators and two or three $\pi$-$\pi$ operators in the variational basis. We extract the lowest three or four energy states using the variational method~\cite{Luscher:1990ck}. For each energy we compute the associated phase-shift and then extract the resonance parameters using both Breit-Wigner parametrization and a unitary chiral perturbation theory (\uchipt) model based on Ref.~\cite{Oller:1998hw}. 


The outline of this paper is the following: in Section~\ref{sec:method}, we discuss the technical details of our analysis. In Section~\ref{sec:result}, we present the results for the energy spectrum and the associated phase shifts. In Section~\ref{sec:rho}, we discuss the extraction of the resonance parameters and the extrapolation to the physical point. In Section~\ref{sec:conclusion}, we present our conclusion and discuss future plans.

\section{Technical details}
\label{sec:method}

\subsection{Phase-shift formulas}
\label{sec:phaseshift_formula}

As we mentioned in the introduction, L\"{u}scher derived a relation between the two-hadron state energies and their scattering phase-shift~\cite{Luscher:1990ux}. In this study we use extensions of this formula to elongated boxes and to states boosted along the elongated
direction. In this section we collect the relevant formulas.


In a finite volume box with periodic boundary conditions, the internal symmetries such as flavor and isospin are the same as in the continuum. However, the spatial symmetries are reduced to the symmetry group of the box (at least for zero-momentum states). In this study, we consider boxes elongated in one direction, which we take to it be the $z$-direction. For this geometry the rotational symmetry is reduced to the $D_{4h}$  group which is a subgroup of the full rotation group $SO(3)$. In this case, the multiplets transforming under the irreducible representations (irreps) of the $SO(3)$ group are no longer irreducible under the action of $D_{4h}$ group. Instead, they are split into multiplets corresponding to irreps of the $D_{4h}$ group. The resulting split for the lowest angular momentum multiplets is listed in Table~\ref{tab:D4h}. The $\rho$ resonance has angular momentum $l=1$ and negative parity. The irrep $l=1$ will split into $A_2^-$ and $E^-$ irreps. $A_2^-$ is a one-dimensional irrep and the lowest states in this channel correspond to $\pi$-$\pi$ states with a back-to-back momentum along the elongated directions. $E^-$ is a two dimensional irrep and the lowest states in this channel correspond to pions moving in the two transversal directions. These later states change very little when varying the elongation of the box. Since we want to vary the scattering momentum using the elongation of the box, we will focus on states in the $A_2^-$ irrep. 

We note that the states in the $A_2^-$ irrep belong to different irreps of $SO(3)$. From Table~\ref{tab:D4h} we can see that the $A_2^-$ irrep couples not only to $l=1$, but also to other higher angular momentum channels such as $l=3$, $l=5$, and so on. However, to study the $\rho$ resonance, we are interested in two pion states with relatively small scattering momenta. In this energy region the phase-shifts for the $l\ge3$ channels are small and their contribution can be safely neglected.

\begin{table}
\begin{tabular}{c c }
\toprule
$\ell$ & $D_{4h}$ \\
\hline
0 & $A_1^+$\\
1 & $A_2^-\oplus E^-$\\
2 & $A_1^+\oplus B_1^+ \oplus B_2^+ \oplus E^+$\\
3 & $A_2^- \oplus B_1^- \oplus B_2^- \oplus 2 E^-$\\
4 & $2A_1^+ \oplus A_2^+ \oplus B_1^+ \oplus B_2^+ \oplus 2E^+$\\
\bottomrule
\end{tabular}
\caption{Resolution of angular momentum from irrep of $D_{4h}$ group}
\label{tab:D4h}
\end{table}

L\"{u}scher's formula for zero-momentum states in asymmetric boxes was derived previously~\cite{Feng:2004ua} and the possibility to use elongated boxes to scan resonances was also considered in Ref.~\cite{Doring:2011vk}. We present here the form for elongated boxes, with geometry $L\times L\times \eta L$, we used in our previous study~\cite{Pelissier:2012pi,Pelissier:2011ib}. We use the generalized zeta function
\beq
\mathcal{Z}_{lm} (s,q^2;\eta) = \sum_{n\in Z^3} \frac{\mathcal{Y}_{lm}(\tilde{\bm n})}{(\tilde{\bm n}^2-q^2)^s}
\label{eq:zeta_func_rest}
\eeq
where the harmonic polynomials are 
\beq
\mathcal{Y}_{lm}(\tilde{\bm n})= \tilde{n}^l Y_{lm}(\Omega_{\tilde{\bm n}})
\eeq
with 
\beq
\tilde{\bm n} = (n_1,n_2,n_3/\eta), \bm n\in \mathds{Z}^3\,. 
\eeq
The phase shift formula relevant for the $A_2^-$ irrep of the $D_{4h}$ group is
\beq
 \cot\delta_1(k)=\mathcal{W}_{00}+\frac{2}{\sqrt{5}}\mathcal{W}_{20},
\label{eq:phaseshift_formula}
\eeq
where the $\mathcal{W}$ function is 
\beq
\mathcal{W}_{lm}(1,q^2;\eta) = \frac{\mathcal{Z}_{lm}(1,q^2;\eta)}{\pi^{3/2}\eta q^{l+1}} \,.
\eeq
The normalized pion momentum $q$ is defined in terms of the pion momentum $k$,
\beq
q =\frac{kL}{2\pi}\quad\text{with}\quad E =  2 \sqrt{m_{\pi}^2 + k^2}\,,
\eeq
where $E$ is the energy of the two-pion state and $m_\pi$ is the pion mass.

For boosted states with total momentum $\bm P$, the relativistic effects contract the box along the boost direction~\cite{Rummukainen:1995vs}. In the case of an elongated box a boost in a generic direction will further reduce the symmetry group from $D_{4h}$ to a subgroup which depends on the direction of the boost. In this study, we consider states that have a non-zero momentum parallel with the elongated direction. In this case, the length contraction affects only the elongated direction. Therefore the boost does not change the rotational symmetry group which is still $D_{4h}$. As a result, we can still focus on the $A_2^-$ irrep and use the same phase shift formula as in~Eq.~\ref{eq:phaseshift_formula} with a slight modification. 

For the boosted states in a cubic box with momentum ${\bm P}= (2\pi/L) {\bm d}$, where $\bm d$ is a triplet of integers,  the relevant zeta function is
\beq
\mathcal{Z}_{lm}^{\bm d}(s;q,\gamma) = \sum_{\bm n\in P_{\bm d}(\gamma)} \frac{\mathcal{Y}_{lm}(\bm n)}{(\bm n^2-q^2)^s},
\label{eq:zeta_func_boost}
\eeq
where 
\beq
P_d(\gamma)=\left\{\bm n\in\reals^3 \mid \bm n=\hat{\gamma}^{-1}\left(\bm m+\frac{\bm d}{2}\right),\bm m\in \mathds{Z}^3\right\}.
\eeq
The projector $\hat{\gamma}^{-1}$ is defined as 
\beq
\hat{\gamma}^{-1} \bm n = \bm n_{\parallel}/\gamma+\bm n_{\perp}, \bm n_{\parallel} = \bm v(\bm n\cdot\bm v)/\bm v^2  \ \text{and}\ \bm n_{\perp}=\bm n - \bm n_{\parallel}.
\label{eq:gamma}
\eeq
The Lorentz boost factor $\gamma$ can be obtained from the velocity of the boost: $\gamma= 1/\sqrt{1-\bm v^2}$, where $\bm v= \bm P/E$. The energy in the center-of-mass frame is related the energy in the lab frame
\beq
E_{\text{CM}} = E/\gamma.
\label{eq:energy}
\eeq
The phase shift formula is the same as in Eq.~\ref{eq:phaseshift_formula} but with a modified $\mathcal{W}$
\beq
\mathcal{W}_{lm}(1,q^2;\gamma)= \frac{\mathcal{Z}^{\bm d}_{lm}(1,q^2;\gamma)}{\pi^{\frac{3}{2}}\gamma q^{l+1}}\,.
\label{eq:w_function}
\eeq

We extend now the phase shift formula to boosted states in an elongated box, with the boost in the elongated direction. The only effect of the elongation is that the summation region $P_{\bm d}(\gamma)$ changes to
\beq
P_{\bm d}(\gamma,\eta) =\left\{\bm n\in\reals^3 \mid \bm n=\hat{\gamma}^{-1} \hat{\eta}^{-1}(\bm m+\frac{1}{2}\bm d), \bm m\in \mathds{Z}^3 \right\} \,,
\label{eq:eta}
\eeq
with $\hat{\eta}^{-1}\bm m =(m_x,m_y,m_z/\eta)$, assuming that the boost and elongation are in the $z$-direction.

To evaluate the phase shift formula in~Eq.~\ref{eq:phaseshift_formula}, we need to compute $\mathcal{Z}_{00}^{\bm d}(s=1)$ and $\mathcal{Z}_{20}^{\bm d}(s=1)$. The zeta functions as defined in~Eq.~\ref{eq:zeta_func_rest} and Eq.~\ref{eq:zeta_func_boost} diverge at $s=1$ and we need to compute them via an analytical continuation. The details for evaluating these functions are presented in the Appendix~\ref{appendix:zeta_func}.

\subsection{Interpolating basis}
\label{variational_method}

In order to extract several low-lying energy levels from the Euclidean correlation functions, 
we use the variational method proposed by L\"{u}scher and Wolff~\cite{Luscher:1990ck}. 
The idea is to construct a correlation matrix using a set of interpolating fields with the 
same quantum numbers and extract the energy levels by solving 
an eigenvalue problem. Choosing a set of interpolating fields with different couplings to the
the eigenstates of the Hamiltonian helps resolve energy states that are nearly degenerated. 
In our case the interpolating field set will
include both quark-antiquark (single-hadron) and multi-hadron interpolating fields. 

The correlation matrix is constructed from two-point functions of all the 
interpolating fields in the basis. If we denote the $N$ interpolators in the basis with
${\mathcal{O}_i}$ with $i=1,\ldots,N$, the elements of the correlation matrix are
\beq
C_{ij}(t)=\av{\mathcal{O}_i(t) \mathcal{O}_j^\dagger(0)} \,.
\eeq
We compute the eigenvalues of the correlation matrix by solving the generalized 
eigenvalue problem
\beq
C(t_0)^{-\frac{1}{2}}C(t)C(t_0)^{-\frac{1}{2}}\psi^{(n)}(t,t_0)=\lambda^{(n)}(t,t_0)\psi^{(n)}(t,t_0)
\label{eq:corr}
\eeq
for a particular initial time $t_0$ and for each time slice $t$. For $t\in[t_0, 2t_0]$ the eigenvalues were shown to behave as~\cite{Luscher:1990ck,Blossier:2009kd}
\beq
\lambda^{(n)}(t,t_0)\propto e^{-E_n t}[1+\mathcal{O}(e^{-\Delta E_n t})], \quad n=1,\ldots,N\,,
\eeq
where the correction is driven by the energy difference 
$\Delta E_n= E_{N+1}-E_n$. This long-time behavior shows that the larger interpolating basis 
we use, the faster the correction for the low energy states vanishes. However, since the
energy eigenstates get denser in the higher-energy part of the spectrum, the payoff of the variational method decreases as the size of the correlation matrix increases. 

As we explained earlier we focus on the states in the $A_2^-$ irrep, mainly because the lowest states in this channel correspond to scattering states where the pions move in the elongated direction. The energy of these states changes as we increase the elongation and we can scan the resonance region. For the volumes considered in this study, the elastic region, $E<4m_\pi$, contains only the lowest three or four states and our focus will be on designing a set of interpolators that allows us to compute the energies of these states accurately. Note that as the pion mass becomes lower and the volume is increased there are more multi-hadron states in the elastic scattering energy region and the basis would need to be adjusted accordingly.

To extract these states we need a basis that overlaps both with the resonance state, which is expected to have mainly a quark-antiquark content, and also with the states that have a dominant two-pion content. From a numerical point of view, the quark-antiquark interpolators are advantageous, since they lead after Wick contraction to two-point quark-correlation functions which can be evaluated cheaply using lattice QCD techniques. The four quark-antiquark interpolators are of the form
\beq
\rho^0(\Gamma_i(\bm p), t) = \frac1{\sqrt2} [ \bar{u}(t) \Gamma_i(\bm p) u(t) - \bar{d}(t) \Gamma_i(\bm p) d(t) ] \,.
\eeq
Here we consider $u(t)$ and $d(t)$ to be the quark field on the entire $t$ time slice, a column vector of size 
$N=12\times N_x\times N_y \times N_z$, and $\Gamma_i(\bm p)$ to be $N\times N$ matrices. To help with notation we define $\Gamma_i'(\bm p)$
using
\beq
[\rho^0(\Gamma_i(\bm p), t)]^\dagger = \rho^0(\Gamma_i'(\bm p),t) \,.
\eeq
The structure of $\Gamma_i(\bm p)$ for the quark-antiquark interpolators is listed in the first four
rows of Table~\ref{tab:interp}. 
Two of the interpolators are point-like and differ only in the gamma-matrix structure and the other two involve a covariant derivative
\beq
(\nabla_k)_{x,y}^{ab} = U_k^{ab}(x) \delta_{x+\hat k,y} - \delta^{ab}\delta_{x,y} \,,
\eeq
and they involve quark-antiquark pairs separated by one lattice spacing. 

{\renewcommand{\arraystretch}{1.2}
\begin{table}[b]
\center
\begin{tabular}{*{5}{>{$}r<{$}}}
\toprule
i &\phantom{abc}& \Gamma_i(\bm p) &\phantom{ab}& \Gamma_i'(\bm p) \\
\midrule
 1 && \gamma_3 e^{i \bm p} && \gamma_3 e^{-i \bm p} \\
 2 && \gamma_4\gamma_3 e^{i\bm p} && \gamma_4\gamma_3 e^{-i\bm p}\\
 3 && \gamma_3 \nabla_j e^{i\bm p} \nabla_j && -\gamma_3 \nabla_j e^{-i\bm p} \nabla_j \\
 4 && \frac12 \{e^{i\bm p}, \nabla_3\} && -\frac12\{ e^{-i\bm p}, \nabla_3\} \\
 \rule{0pt}{3.25ex} 
 5 && \gamma_5 e^{i\bm p} && -\gamma_5 e^{-i\bm p} \\
\bottomrule
\end{tabular}
\caption{Interpolator structure for the quark bilinears used in this study. The first four rows are 
used for the quark-antiquark interpolators and the last row is used for the pion-pion interpolators.
The elongation is assumed to be in the $z$-direction and the interpolators are chosen so that the 
$\rho$ polarization is longitudinal: $\gamma_3$ and $\nabla_3$ need to be changed accordingly
if the elongation direction is changed.}
\label{tab:interp}
\end{table}
}

Unfortunately, the quark-antiquark interpolators overlap very poorly with the multi-hadron state (the overlap is suppressed by a power of the lattice volume~\cite{Dudek:2012xn}). Therefore we have to include also pion-pion interpolators in our basis.
The pion-pion interpolators are constructed to have isospin $I=1$ and $I_3=0$, corresponding to $\rho^0$:
\beq
\pi\pi(\bm p_1,\bm p_2)= \frac{1}{\sqrt{2}} \{ \pi^+(\bm p_1)\pi^-(\bm p_2)-\pi^-(\bm p_1)\pi^+(\bm p_2)\}\,.
\label{eq_pipi_op_1}
\eeq
Here we use
\beqs
\pi^-(\bm p,t) &= \sum_{\bm x} \bar{u}(\bm x,t) \gamma_5 d(\bm x,t) e^{i\bm p\bm x}=\bar{u}(t) \Gamma_5(\bm p) d(t) \,, \\
\pi^+(\bm p,t) &= \sum_{\bm x} \bar{d}(\bm x,t) \gamma_5 u(\bm x,t) e^{i\bm p\bm x}=\bar{d}(t) \Gamma_5(\bm p) u(t) \,.
\eeqs

To construct interpolators transforming according to $A_2^-$ representation, we can start with any interpolator that has some $A_2^-$ 
component and project onto the relevant subspace:
\beq
\pi\pi(\bm p_1,\bm p_2)_{A_2^-} = \frac1{|D_{4h}|} \sum_{g\in D_{4h}} \chi_{A_2^-}(g) \pi\pi(R(g) \bm p_1, R(g) \bm p_2) \,,
\eeq
where $R(g)$ implements the rotation associated with the symmetry transformation $g$, and $\chi_{A_2^-}$ is the character of $g$ in
the $A_2^-$ irrep. 

For states with zero total momentum, $\bm P_0=\bm p_1 + \bm p_2 = 0$ and for non-zero momentum states with
$\bm P_1 = (0,0,1)$ we use the following interpolators
\beqs
\pi\pi_{001}^{(i)} &= \pi\pi( \bm p_1=(0,0,1), \bm p_2=\bm P_i-\bm p_1 ) \,,\\
\pi\pi_{002}^{(i)} &= \pi\pi( \bm p_1=(0,0,2), \bm p_2=\bm P_i-\bm p_1 ) \,,\\
\pi\pi_{011}^{(i)} &= \frac12 \sum_{p_1\in {\cal P}}\pi\pi( \bm p_1, \bm p_2=\bm P_i-\bm p_1 ) \,,\\
\eeqs
where ${\cal P}=\{(0,1,1),(1,0,1),(-1,0,1),(0,-1,1)\}$, is the set of momenta generated by symmetry transformations 
$R(g)\bm p$ from $\bm p= (0,1,1)$ which have $p_z>0$. The later condition is imposed for different reasons for the $\pi\pi^{(0)}$
and $\pi\pi^{(1)}$ interpolators. In the zero momentum case we impose it because the interpolators 
$\pi\pi( \bm p, -\bm p)$ and $-\pi\pi( -\bm p, \bm p)$ are identical. For non-zero momentum, the symmetry group
transformations mix states with different total momentum, $\bm P_1$ and $-\bm P_1$. When computing the correlation functions 
of such interpolating fields, the correlation functions between sink and source of different momentum vanish. The non-vanishing 
contributions connect states with the same total momentum. The expectation values for correlations functions associated with 
momentum $\bm P_1$ and $-\bm P_1$ are the same due to symmetry, so we only need to evaluate the contributions due to momentum $\bm P_1$.

The same interpolators for the non-zero momentum case can also be derived using an analysis based on symmetries of the 
Poincare group on the lattice~\cite{Moore:2005dw}. In our case the little group for states with momentum $\bm P_1$ is $C_{4v}$ and 
the relevant irrep is $A_1$ since the longitudinal states have projection $0^+$ in the momentum direction. We prefer to derive them
from projections onto the $A_2^{-}$ irrep of $D_{4h}$ to make clear that the connection between energies and phase-shifts is provided
by the relation in Eq.~\ref{eq:phaseshift_formula}.

To summarize, we use four quark-antiquark interpolators and two pion-pion interpolators for most ensembles to 
form a $6\times6$ variational basis. For the ensembles with the largest elongation, 
$N_z=48$ for the largest pion mass and $N_z=32$ for the lowest mass, we add a third pion-pion interpolator $\pi\pi_{002}^{(i)}$, 
for reasons that will be explained later. In principle, six or seven energies can be extracted from the correlation matrix. 
However, we only focus on the first three lowest energy levels that are located in the elastic scattering region with better 
signal-to-noise ratio. 

\subsection{LapH correlation functions}

Our interpolator basis has quark-antiquark and pion-pion operators. The correlation matrix will have then
three types of entries
\beqs
C&_{\rho_i\leftarrow\rho_j} = \av{\rho_i(\bm P, t_f) \rho_j^\dagger(\bm P, t_i)}=
-\av{
[i\bm P f|j'\bm P i]
} \,,
\\
C&_{\rho_i\leftarrow\pi\pi}=
\av{\rho_i(\bm P,t_f) \pi\pi(\bm p, \bm P-\bm p, t_i)^\dagger} \\
&= \av{[i\bm P f|5'\bm P-\bm p i|5'\bm p i]-[i\bm P f|5'\bm p i|5'\bm P-\bm p i]} \,, \\
C&_{\pi\pi\leftarrow\pi\pi}=\av{\pi\pi(\bm p',\bm P-\bm p',t_f)\pi\pi(\bm p,\bm P-\bm p,t_i)^\dagger}\\
&=\big\langle
[5\bm p'f|5\bm P-\bm p'f|5'\bm p i|5'\bm P-\bm p i] \\
&-[5\bm p'f|5\bm P-\bm p'f|5'\bm P-\bm p i|5'\bm p i] \\
&+[5\bm P-\bm p'f|5\bm p'f|5'\bm P-\bm pi|5'\bm pi] \\
&-[5\bm P-\bm p'f|5\bm p'f|5'\bm pi|5'\bm P-\bm pi] \\
&+[5\bm p'f|5'\bm pi][5\bm P-\bm p'f|5'\bm P-\bm pi] \\
&-[5\bm p'f|5'\bm P-\bm pi][5\bm P-\bm p'f|5'\bm pi]
\big\rangle \,.
\label{eq:contractions}
\eeqs
Above we used the following notation for the traces produced by Wick contractions
\beq
[i_1\bm p_1 j_1|\ldots|i_k \bm p_k j_k]\equiv 
\Tr\prod_{\alpha=1}^k \Gamma_{i_\alpha}(\bm p_\alpha) M^{-1}(t_{j_\alpha},t_{j_{\alpha+1}}) \,,
\label{eq:trace}
\eeq
where $j_{k+1}$ is defined to be $j_1$ and $M^{-1}(t,t')=\av{u(t)\bar{u}(t')}$ is the quark propagator
between time slices $t$ and $t'$, viewed as a $N\times N$ matrix (for more details about the notation
see~\cite{Pelissier:2012pi}). Note that $\Gamma_{i'}(\bm p)$ is meant to be $\Gamma_i'(\bm p)$. We also
note that when $\bm P=0$, the number of diagrams that needs to be evaluated is reduced to one for three
point functions and to four for four-point functions.

The two-point quark diagrams can be evaluated cheaply by computing the quark propagator from one
point on the lattice and using the translational invariance. This is not possible for three and four-point
diagrams. In this case the all-to-all propagator needs to be computed which is not practical. 
The LapH method~\cite{Peardon:2009gh,Morningstar:2011ka} offers a way to address this problem.
This method can be understood as a form of smearing of the quark fields, both at the source and sink,
with the added bonus that the calculation can be completely expressed in terms of a quark-propagator
reduced to a subspace of slowly moving quark states. The smearing is purely in the spatial direction and
it is gauge covariant by construction. As such the smeared quark fields have the same transformation
properties under lattice symmetry transformations and interpolators built out of smeared fields
have the same quantum numbers as the ones built using the original fields.

The smearing is constructed using the eigenvectors of the three-dimensional covariant Laplace operator,
\beq
\Delta_t = -\sum_{k=1}^3 \nabla_k(t)\nabla_k(t)^\dagger\,,
\eeq
with the components
\beqs
\Delta_t^{ab}(\bm x,\bm y)& =
\sum^3_{k=1}[U_k^{ab}(\bm x,t) \delta(\bm x+\hat{k},\bm y) \\
     & +U_k^{ba}(\bm y,t)^*\delta(\bm x-\hat{k},\bm y)-2\delta^{ab}\delta(\bm x,\bm y)] \,.
\eeqs
This operator is negative-definite and its eigenvalues are all negative. We sort the eigenvalues so that 
$\lambda_1(t)>\lambda_2(t)>\ldots$. Using the eigenvectors $\ket{s;t}$ of $\Delta_t$ corresponding to 
eigenvalue $\lambda_s(t)$, we define the smearing operator:
\beq
\mathcal{S}(t) \equiv \sum_{s=1}^{N_v} \ket{s;t} \bra{s;t}\,,
\eeq
which is the projector on the space spanned by the $N_v$ lowest frequency eigenmodes of the Laplacean operator.
The smearing operator only acts on the spatial and color space. The smeared quark field is
\beq
\tilde{q}(t) \equiv \mathcal{S}(t) q(t)\,.
\eeq
As mentioned earlier, the bilinears $\tilde{\bar{q}}\Gamma\tilde{q}$ have the same transformation properties as
$\bar{q}\Gamma q$ and we can use them as building blocks for the $\rho$ and $\pi\pi$ interpolators defined in the 
previous section. The advantage of this substitution is that, on one hand, the correlation functions will be less noisy,
since the overlap of these interpolators with the physical states is better when we choose $N_v$ appropriately.
On the other hand, as we will show below, the calculation of all the correlation functions requires only the evaluation
of the quark propagators from $4\times N_v\times N_t$ sources, which is a significant improvement over evaluating the
all-to-all propagator when $4 N_v\ll N$.

After Wick contractions, the correlation functions are identical in form with the ones in Eq.~\ref{eq:contractions}, but the
propagator that appears in the spinorial traces in Eq.~\ref{eq:trace} is replaced with a smeared version
\beq
\av{q(t)\bar{q}(t')} \rightarrow \av{\tilde{q}(t)\tilde{\bar{q}}(t')}
=\mathcal{S}(t)M^{-1}(t,t')\mathcal{S}(t')   \,.
\eeq
The traces in Eq.~\ref{eq:trace} are then replaced with a smeared version
\beqs
\bb i_1\bm p_1 j_1|&\ldots|i_k \bm p_k j_k \rr \equiv \\
&= \Tr\prod_{\alpha=1}^k \Gamma_{i_\alpha}(\bm p_\alpha) S(t_{j_\alpha})M^{-1}(t_{j_\alpha},t_{j_{\alpha+1}})S(t_{j_{\alpha+1}}) \\ 
&=\Tr\prod_{\alpha=1}^k \tilde{\Gamma}_{i_\alpha}(\bm p_\alpha,t_{j_\alpha}) \tilde{M}^{-1}(t_{j_\alpha},t_{j_{\alpha+1}}) \,,
\eeqs
where
\beqs
\tilde{\Gamma}_i(\bm p,t)_{s,s'}^{\alpha\beta} &\equiv  \opbraket{s;t}{\Gamma_i(\bm p)^{\alpha\beta}}{s';t} \,,\\
\tilde{M}^{-1}(t,t')_{s,s'}^{\alpha\beta} &\equiv \opbraket{s;t}{M^{-1}(t,t')^{\alpha\beta}}{s';t'}
\eeqs
are $(4 N_v)\times(4 N_v)$ matrices. Above, $s, s'$ are eigenvector indices and $\alpha, \beta$ are spinorial indices.
These relations can be easily derived using the definition of the smearing operator and
the cyclic property of the trace. We note then that the we only require the evaluation of the smeared all-to-all propagator
$\tilde{M}^{-1}$ which only requires $4 N_v\times N_t$ inversions compared to $N\times N_t$ for the all-to-all
propagator. For example, even on the smallest lattice used in this study $N=12\times 24^3=165,888$ whereas $4N_v=400$, a significant
reduction.

We also note that the traces are over matrix products with dimensions $4N_v$. When evaluating a large number of diagrams, the
bottleneck becomes the matrix-matrix products. It is then advisable to carefully examine the required products to reduce the
calculation. One such simplification can be implemented for matrices in this $4N_v$ space that factorize in a tensor product
between the spinorial and Laplacean subspaces. For example
\beq
\tilde \Gamma_1(\bm p;t) = \gamma_3 \tilde{e}^{i\bm p}(t)\,,
\tilde{e}^{i\bm p}(t)_{s,s'} \equiv \opbraket{s;t}{e^{i\bm p}}{s';t}\,.
\eeq
The multiplication with this matrix can be implemented four times more efficient than when using a full $(4N_v)\times(4N_v)$
representation for the $\tilde{\Gamma}_1$ matrix.

\begin{figure}[t]
\includegraphics[scale=0.6]{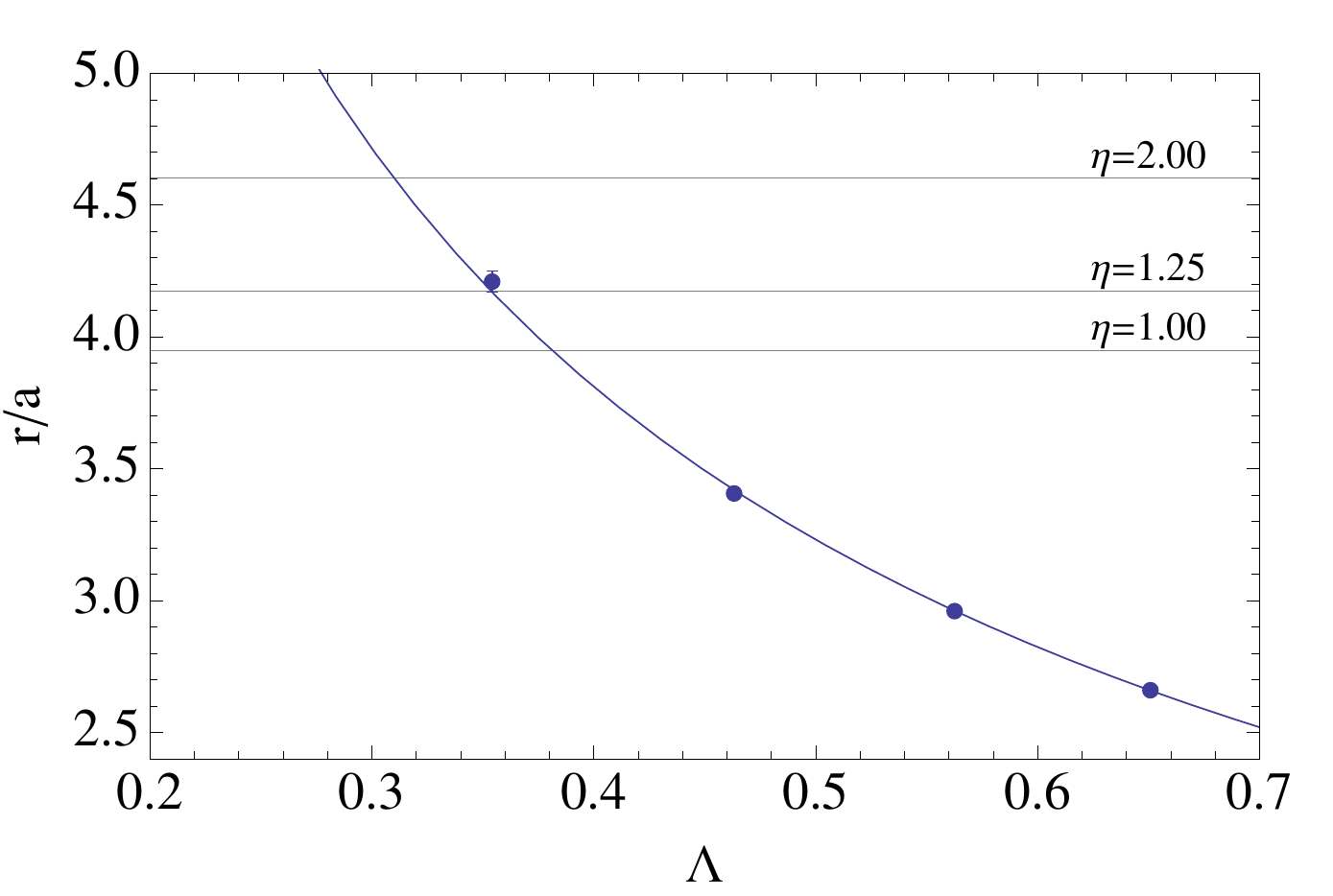}
\caption{Smearing radius for $m_\pi=315\MeV$ and lattice spacing $a\approx0.121\fm$.
The radius is evaluated on a $16^3\times 32$ ensembles for $N_v=25$, 50, 75, and 100 and the line represents a power law fit. 
The horizontal lines indicate the smearing radius corresponding to $N_v=100$ on ensembles $\mathcal{E}_1$, $\mathcal{E}_2$, 
and $\mathcal{E}_3$.}
\label{fig:smear_radius}
\end{figure}

The action of the smearing operator $\mathcal{S}$ can be illustrated by acting on a point source. The magnitude of 
$\norm{(\mathcal{S}\delta_{\bm x})(\bm y)}$ decays like a gaussian away from the source, $\exp(-\norm{\bm x-\bm y}^2/r^2)$. 
The smearing radius $r$ depends of the on the energy cutoff, $\Lambda=-\lambda_{N_v}(t)$. 
We can determine the optimal $\Lambda$ by tuning individual operators to minimize the errorbars of the effective mass at a
fixed time~\cite{Peardon:2009gh}. In this study, we fix the number of Laplacean eigenvectors to $N_v=100$. Since the
volume varies with the ensemble, the energy cutoff and the smearing radius change with the ensemble too. In Fig.~\ref{fig:smear_radius} 
we plot the smearing radius as a function of $\Lambda$ for the higher mass ensembles. We indicate in the figure the smearing radii
for $\eta=1$, $1.25$, and $2$. Note that the change in the smearing radius from the smallest to the largest volume is about 10\%,
so the smearing is very similar on all ensembles, with $r\approx 0.5\fm$.

Before we conclude we want to make a couple of points about LapH smearing. One benefit of this method is that we can separate the 
calculation of the smeared quark propagator $\tilde{M}^{-1}$ from computing the correlation functions. This is very important when 
using a large variational basis, especially since it allows to add other interpolating fields to the basis without
having to redo the inversions. Another point that we want to stress is that the smearing employed here does not
represent an approximation. The smeared interpolating fields have the right symmetry properties even when the number of
Laplacean eigenvectors $N_v$ is very small. If the number is too small the overlap with the physical states is poor and
the signal-to-noise ratio will be bad. Finally, even though the number of inversions is much smaller than the total number
required for the all-to-all propagator, we still need to compute $4N_v\times N_t$ inversions for each configuration: 
19,200 and $25,600$ inversions per configuration for the $m_\pi=315\MeV$ and $m_\pi=226\MeV$ ensembles, respectively. 
This calculation was done using a GPU implementation of a BiCGstab inverter~\cite{Alexandru:2011ee}.

\subsection{Fitting method}
\label{sec:fitting_method}

To extract the mass and width of the resonance we need
to fit the phase-shift data using a phase-shift parametrization
in the resonance region. For the $\rho$-resonance a Breit-Wigner
parametrization
\beq
\label{eq:bw-param}
\tan \delta(E) = \frac{E\, \Gamma(E)}{m_\rho^2-E^2}
\quad \text{with} \quad \Gamma(E) = \frac{g_{\rho\pi\pi}^2}{6\pi} 
\frac{p^3}{E^2},
\eeq
describes the phase shift well close to the resonance. For
a given box geometry, this parametrization can be used to 
determine the eigenvalues of the Hamiltonian using L\"uscher's
formula for $A_2^-$ irrep in Eq.~\ref{eq:phaseshift_formula}.

\begin{table*}[t]
\begin{tabular}{@{}*{13}{>{$}c<{$}}@{}}
\toprule
\text{ensemble}& N_t\times N_{x,y}^2\times N_z & \eta & a[\fm]   & N_\text{cfg} & N_v & am_{\pi} & am_K & am_{N} &am^{pcac}_{u/d}  & af_{\pi}  & am^{pcac}_{s} &  af_{K}  \\
\midrule
\mathcal{E}_1&48\times24^2\times24  &  1.0  & 0.1210(2)(24) & 300   & 100 & 0.1934(5) &  0.3235(6) & 0.644(6) & 0.01237(9)& 0.0648(8) &  0.03566(6)        & 0.1015(2)     \\ 
\mathcal{E}_2&48\times24^2\times30  &  1.25 & -        & -     & -   &    -      &     -      &      -   &     -     &  -        &
-          &     -         \\
\mathcal{E}_3&48\times24^2\times48  &  2.0  & -        & -     & -   &    -      &     -      &      -   &     -     &  -        &
-          &     -         \\
\mathcal{E}_4&64\times24^2\times24  &  1.0  & 0.1215(3)(24)& 400   & 100 & 0.1390(5)  &  0.3124(8) & 0.62(1)& 0.00617(9)& 
0.060(1) & 0.03358(6)        & 0.0980(2)   \\
\mathcal{E}_5&64\times24^2\times28  &  1.17 &      -   & -     & -   &    -      &     -      &      -   &   -       &   -       &
-          &    -          \\
\mathcal{E}_6&64\times24^2\times32  &  1.33 &      -   & -     & -   &    -      &     -      &      -   &   -       &   -       &
-          &    -          \\
\bottomrule
\end{tabular}
\caption{The parameters for the ensembles used in this study. The lattice spacing $a$ for each ensemble is listed as 
well as the number of gauge configurations $N_{\text{cfg}}$ and the number of eigenvectors used for LapH smearing. 
$am_{N}$, $af_{\pi}$, and $af_{K}$ represent the nucleon mass, pion decay constant and kaon decay constant in lattice units.
The two errors for the lattice spacing are stochastic, from the $w_0/a$ determination, and a systematic one estimated to be 2\%.}
\label{Tab:ensembles}
\end{table*}

The energies $E_k$ satisfying both equations Eq.~\ref{eq:bw-param} and 
Eq.~\ref{eq:phaseshift_formula} are the expected eigenvalues of the Hamiltonian on
periodic boxes with geometry $L^2\times \eta L$.
These solutions are functions of the geometry of the box $\eta$ and the
parameters of the Breit-Wigner curve and 
we will denote them with $E_k(m_\rho,g_{\rho\pi\pi}; \eta)$.
To determine the fit parameters we minimize the chi-square function
\beq
\chi^2(m_\rho, g_{\rho\pi\pi}) = \sum_{e} \delta_e^T C_e^{-1} \delta_e,
\eeq
where the sum runs over the statistically independent ensembles with 
different elongations and the residue vector is given by 
\beq
(\delta_e)_k = E_k(m_\rho, g_{\rho\pi\pi}; \eta_e) - E^{(e)}_k \,.
\eeq
Above we denote with $E^{(e)}_k$ the $k^\text{th}$ energy extracted from
ensemble $e$ and with $C_e$ the covariance matrix for these energies. Note
that the residue vector includes the residues for both zero-momentum states
and boosted-states and thus it has between 6 and 8 entries depending on the
ensemble considered. The values for $E^{(e)}_k$ and covariance matrix $C_e$
are given in Appendinx~\ref{appendix:fitting}. The energies are extracted
using individual correlated fits and the covariance matrices are estimated
using a jackknife analysis.

\section{Results}
\label{sec:result}

In this section we present the results for the energies and phase-shifts extracted from the ensembles used in this study
and discuss some of the salient issues.
We have generated configurations using L\"{u}scher-Weiss gauge action~\cite{Luscher:1984xn,Alford:1995hw} and nHYP-smeared clover 
fermions~\cite{Hasenfratz:2007rf} with two mass-degenerate quark flavors ($N_f=2$).
For each mass we generated three sets of ensembles with different elongations. The elongations were chosen to ensure that 
the energy spectrum for the zero-momentum states in the $A_2^-$ channel overlaps well with the $\rho$-resonance region,
following the procedure described in a previous study~\cite{Pelissier:2012pi}. 

The parameters for these ensembles are listed
in Table~\ref{Tab:ensembles}. A couple of comments regarding the parameters listed in the table. The lattice spacing
was determined using an observable based on the Wilson flow~\cite{Luscher:2010iy}: the $w_0$ parameter~\cite{Borsanyi:2012zs}.
This quantity can be determined with very little stochastic error from a handful of configurations. We used 150 configurations
from ensembles ${\cal E}_1$ and ${\cal E}_4$ and computed $w_0/a=1.3888(24)$ and $w_0/a=1.4157(37)$ respectively. These measurements
were used to fix the lattice spacing using the conversion factors determined in~\cite{Bruno:2013gha}: we computed the
dimensionless quantity $y=m_\pi^2 t_0$, determined $w_0(y)/w_0(y=0)$ from Fig. 4 in the above reference, and then converted to physical
units using $w_0(y=0)=0.1776(13)\fm$. This value of $w_0$ was determined from a set of $N_f=2$ simulations where $f_K$ was used to set the
scale~\cite{Fritzsch:2012wq}. The scale determined this way differs from the scale we used in our previous study~\cite{Pelissier:2012pi}
by 3.5\%, but we attribute this shift to the fact that the value of the Sommer parameter~\cite{Sommer:1993ce} is difficult
to define unambiguously on configurations with light quarks. The value we used in our previous study was $r_0=0.5\fm$, but 
recent determinations of $r_0$ from global fits of the hadronic spectrum favor smaller values~\cite{Yang:2014sea,Aoki:2010dy}
and produce values in agreement with the scale determined based on $w_0$. 
We decided to adopt the scale determined by $w_0$ because the method is very straightforward and it has small stochastic errors.
Note that at fixed lattice spacing in the presence of lattice artifacts, the lattice spacing determination introduces a systematic 
error. We estimate that our systematic error associated with the lattice spacing is at the level of 2\%.
To confirm the correctness of the lattice spacing we looked at the nucleon mass, pion and kaon decays
constants. We computed the nucleon mass $m_N$ and extrapolated to the physical point using an empirically motivated fit 
form~\cite{WalkerLoud:2008pj}. The extrapolated values agree at the level of 2\%, but this may be fortuitous since the error bars
of the extrapolation were at the level of 4\%. In any case this error level is in line with the expectation from other studies that 
used HEX-smeared fermions at similar lattice spacing~\cite{Durr:2010aw}, where the hadronic spectrum was found to be shifted by 
about 2\% relative to the continuum. The values of $f_\pi$ and $f_K$ were determined using the procedure outlined 
in~\cite{Fritzsch:2012wq}. For the masses used in our study our values for $f_K$ differs by less than 1\% from the values
determined there at much smaller lattice spacing.

\begin{figure}[t]
\includegraphics[scale=0.6]{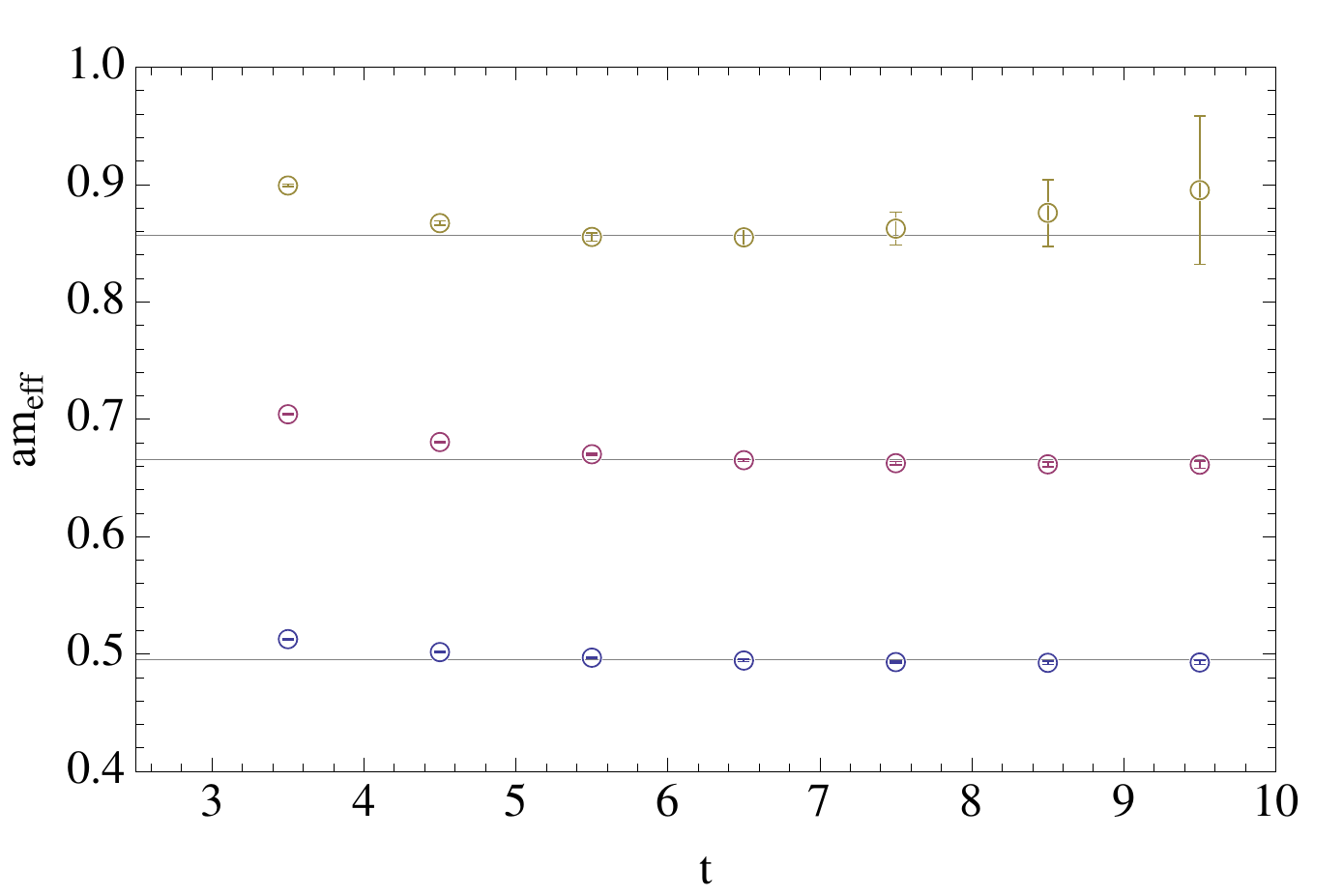}
\centering
\caption{The effective mass for the lowest three energy levels of ${\cal E}_1$ ensemble. 
The dashed line represent the $4 m_{\pi}$ threshold. Contamination from higher energy levels is present in the early time slices, 
therefore we used a double exponential to extract the energy. The horizontal lines indicate the results of the fit.}
\label{Fig:3states}
\end{figure}

For each ensemble we extract the lowest three or four levels in the $A_2^-$ channel, since these levels correspond roughly to the 
elastic region where the center-of-mass energy is below $4m_\pi$. To extract the energies we compute the correlation
matrix $C(t)$, solve the eigenvalue problem in Eq.~\ref{eq:corr}, and fit the extracted eigenvalues to an exponential ansatz. 
In Fig.~\ref{Fig:3states} we show the effective
mass computed from the three lowest eigenvalues on the ${\cal E}_1$ ensemble. Note that the effective mass does not flatten out until
later times. To extract the energy we fit a double exponential function constrained to pass through $1$ at $t=t_0$:  
$f_1(t) = A e^{-E (t-t_0)}+(1-A) e^{-E'(t-t_0)}$. For the lowest $\bm P=0$ state on ensemble ${\cal E}_3$ this fit form does not 
work, due to wrap-around effects in the time direction~\cite{Prelovsek:2008rf,Dudek:2012gj}. We added a constant term to the
fit form to accommodate this effect: $f_2(t)= A e^{-E (t-t_0)}+(1-A-C) e^{-E'(t-t_0)}+C$. We used this fit form with
the other zero-momentum states in all the ensembles, but the constant term produced by minimizing $\chi^2$ was compatible with zero. 
For the moving states, the wrap-around effect leads to a small, slowly decaying term with a rate controlled by the mass difference 
between the moving pion and the pion at rest $\delta E=\sqrt{m_\pi^2 + (2\pi/L)^2}-m_\pi$~\cite{Dudek:2012gj}. 
For the states where this contribution 
was significant, we used the following fit form $f_3(t)=A e^{-E (t-t_0)}+(1-A-C) e^{-E'(t-t_0)}+C e^{-\delta E(t-t_0)}$.
The fitting details including the choice of $t_0$, fitting range, fit form, energy extracted, and quality of the fit are 
tabulated in Table~\ref{Tab:fit_details} in the Appendix.

\begin{figure}[t]
\includegraphics[scale=0.34]{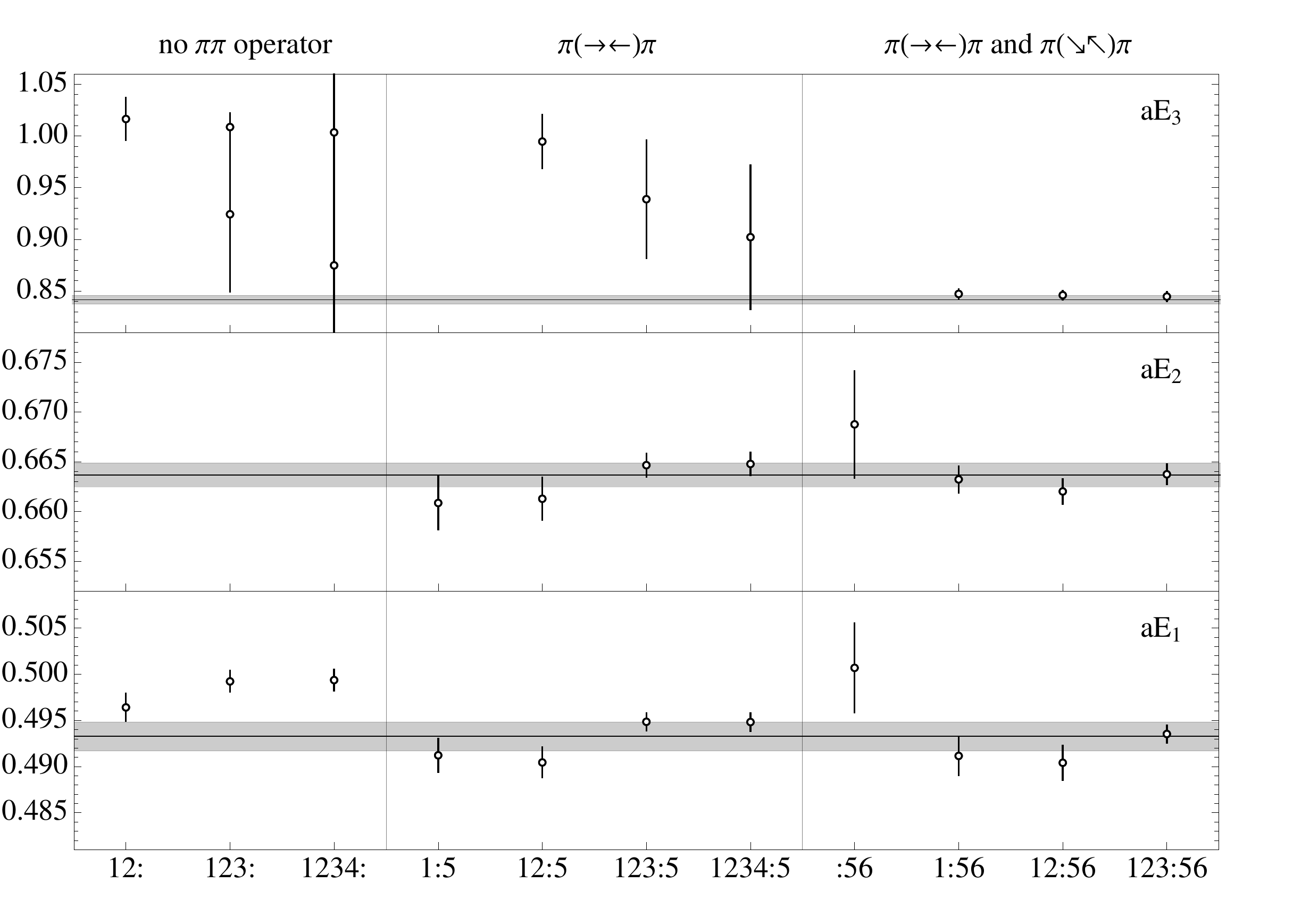}
\caption{Energy spectrum for ensemble ${\cal E}_1$ with different interpolator basis combinations. 
The horizontal axis labels different interpolator choice: $\mathcal{O}_{1-4}$ are the $q\bar{q}$ interpolating fields, 
$\mathcal{O}_{5}$ is $\pi\pi_{100}$ and $\mathcal{O}_6$ is $\pi\pi_{011}$. 
The vertical axis represents the energy for the three lowest levels. Note that we use different scales for each level.
The three horizontal band shows the energy values extracted from $6\times6$ correlation matrix. 
The first panel only includes the $q\bar{q}$ operators. 
Second panel includes $\pi\pi_{100}$ and various $q\bar{q}$ operators. 
The third panel has $\pi\pi_{100},\pi\pi_{011}$ with various $q\bar{q}$ operators.}
\label{Fig:basis_choice}
\end{figure}

We discuss now the choice of the interpolator fields and in particular we address the question whether our interpolating field 
basis overlaps well with the lowest three energy states in the $A_2^-$ channel. To this end, we compare the energy spectrum extracted 
using different subsets of the interpolating fields basis. To simplify the discussion we focus first on the ${\cal E}_1$ ensemble.
The energy spectrum extracted from different interpolating fields basis combinations is plotted in Fig.~\ref{Fig:basis_choice}. 
In the first panel, we include only $q\bar{q}$ operators. While the ground state seems to be well approximated, 
the $q\bar{q}$ operators have little overlap with the first and second excited states which indicates that they are multi-hadron 
states. In the second column, we use the $\pi\pi_{100}$ operator together with various combinations of $q\bar{q}$ operators. 
The ground state and first excited state are well reproduced, even when using only one $q\bar q$ interpolator. However, 
the second excited state has large error-bars even if we add three other $q\bar{q}$ operators, which indicates that it has a large
multi-hadron component. In the third panel, we use two multi-hadron interpolators: $\pi\pi_{100}$ and $\pi\pi_{011}$. 
Once one $q\bar{q}$ operator is added to the basis, all three lowest energy states are well determined with small errorbars. 
Adding more $q\bar{q}$ operators to the basis does not improve the extraction and we conclude that these lowest three states are well
captured by our set of interpolators.

\begin{figure}[b]
\includegraphics[scale=0.4]{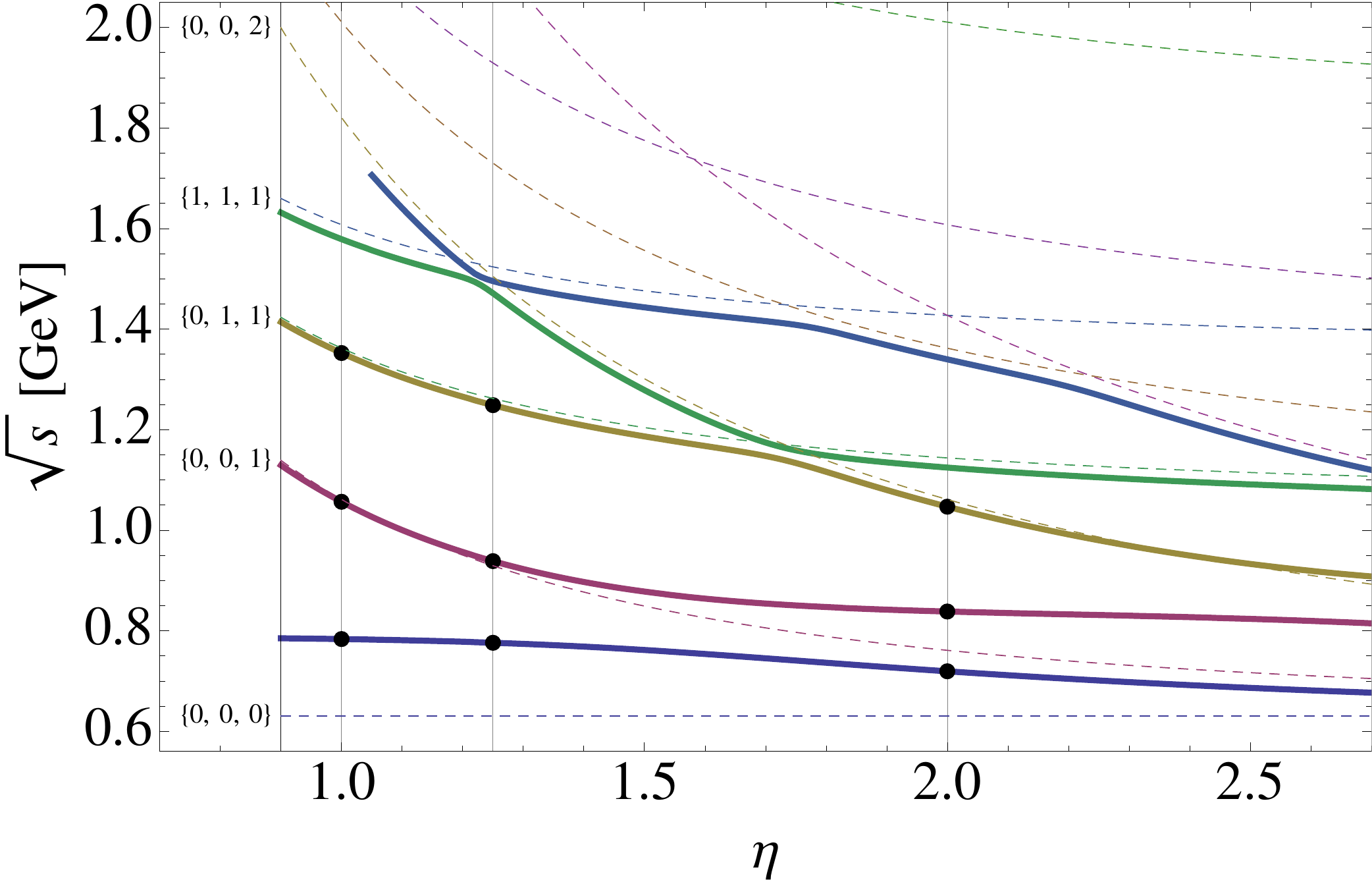}
\caption{Energy spectrum with different elongated factor from unitary chiral perturbation theory in the rest frame $\mathbf{P}=(0,0,0)$~(solid lines). The error bars of eigenvalues are smaller than the symbol size (black circles). $\eta$ labels the elongated factor, in particular $\eta=1.0,1.25,2.0$ are the ensembles we used for $m_\pi\approx315\MeV$. Here, the $\eta=1$ values are fitted and then eigenvalues for $\eta=1.25$ and $\eta=2$ are predicted. The dashed lines represent the energy of non-interacting pion states with various momentum for two pions.}
\label{Fig:expectation}
\end{figure}

\begin{figure*}
\includegraphics[scale=0.60]{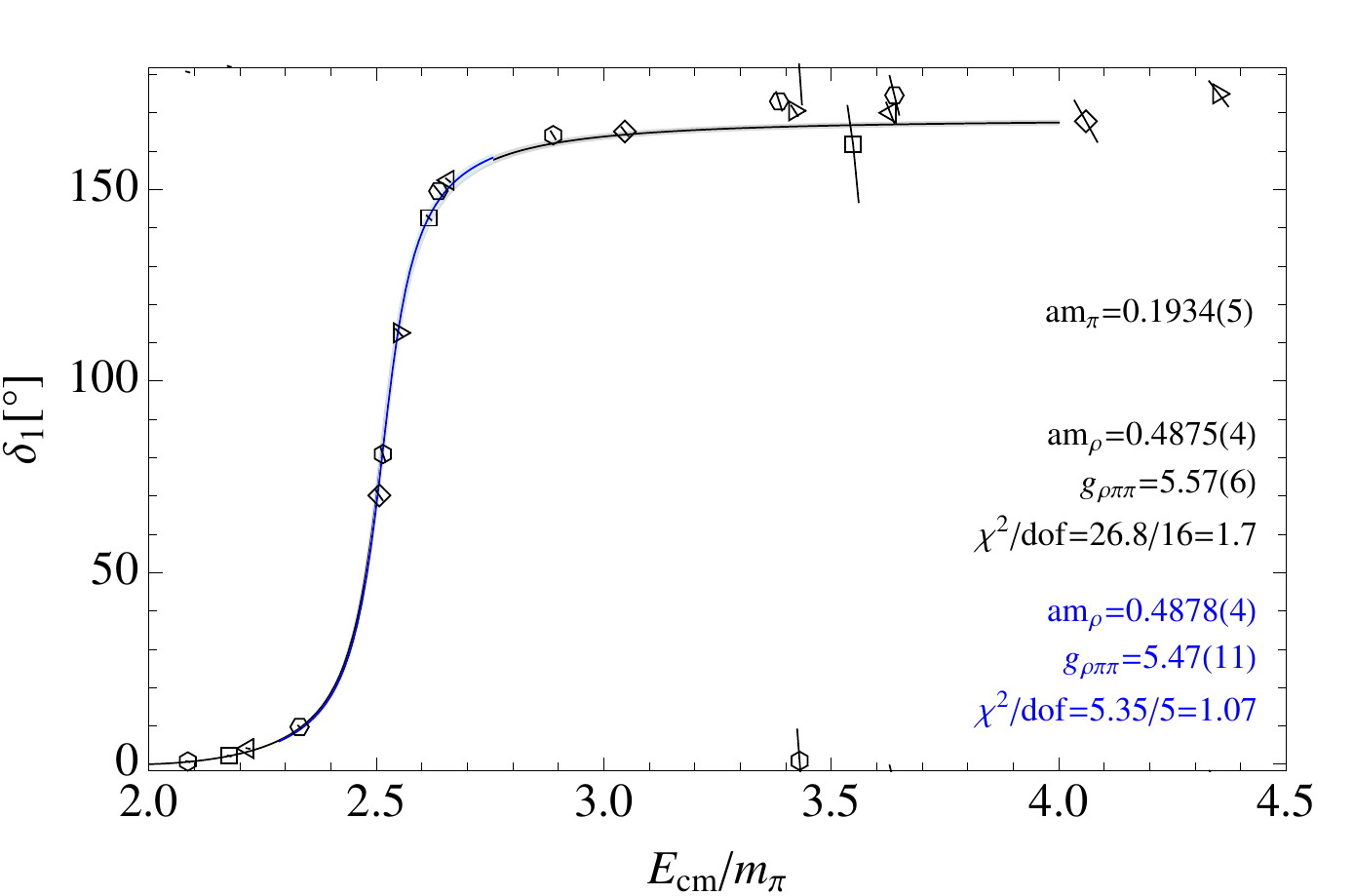}
\includegraphics[scale=0.60]{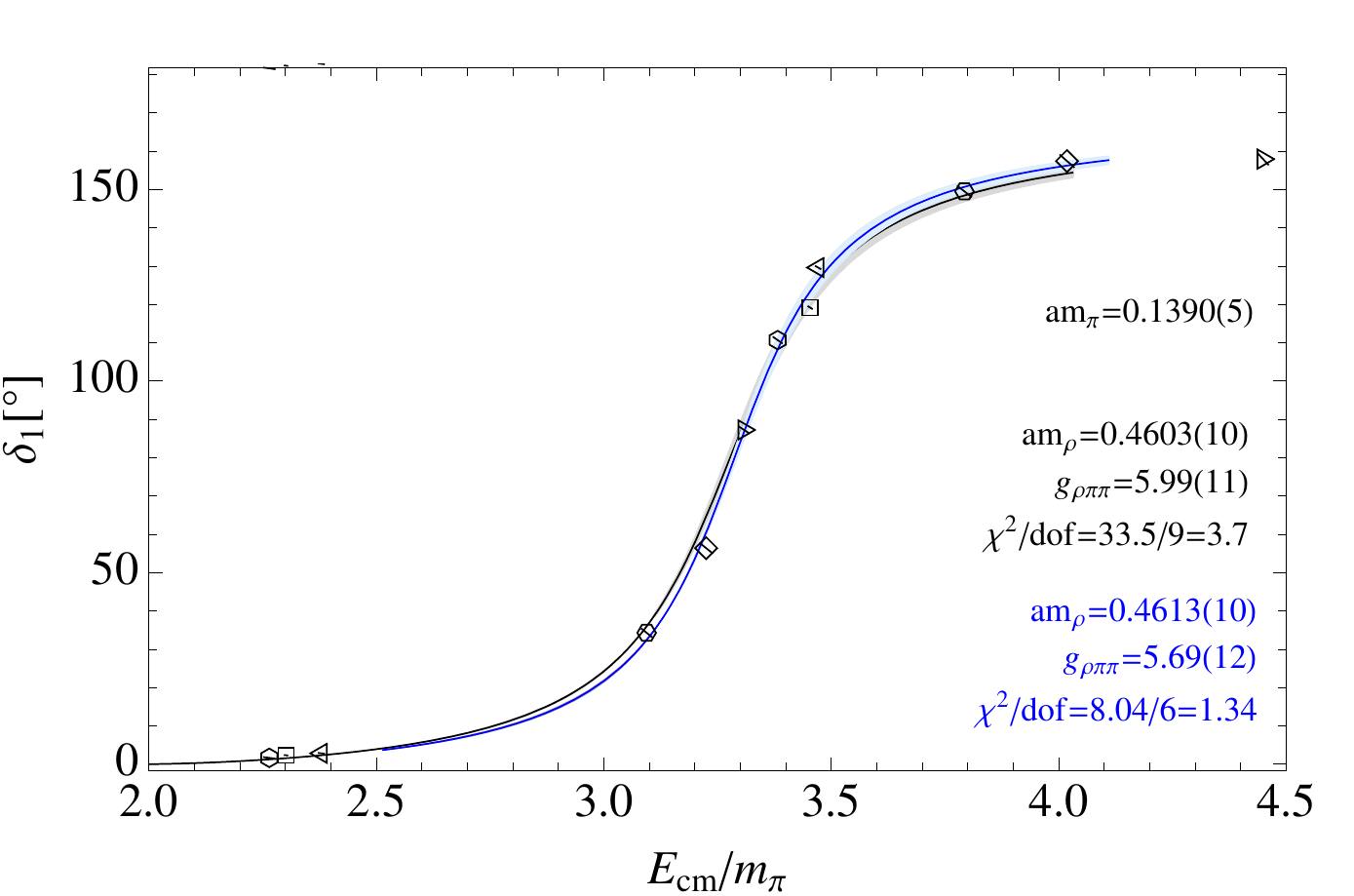}
\caption{Phase shifts as a function of the center of mass energy. The errorbars are slanted along the direction
of the L\"uscher curves. On the left we have the $m_\pi=315\MeV$ data and on the right the $m_\pi=226\MeV$ data.
The triangles, squares, and hexagons correspond to data extracted from ${\cal E}_1$, ${\cal E}_2$, ${\cal E}_3$ (left)
and ${\cal E}_4$, ${\cal E}_5$, ${\cal E}_6$ (right) respectively. The black curve, error-bands, and fit parameters 
correspond to Breit-Wigner fit to all data points in the elastic region, $E_\text{cm}<4m_\pi$. Blue color indicates 
the fit to the data in $m_\rho\pm2\Gamma_\rho$ region. The \uchipt\ fits are very close to the blue Breit-Wigner curves.} 
\label{fig:phase-shift}
\end{figure*}

To confirm the conclusion above, we fit a model based on Unitary Chiral Perturbation theory~(see Appendix~\ref{appendix:uchipt})
to the energy levels extracted from ensemble ${\cal E}_1$ and use it to predict the energy 
levels for different box elongations. The expected energy levels are plotted in 
Fig.~\ref{Fig:expectation} as a function of the elongation. In the graph we also indicate the 
expected energy levels for two-pion states in the absence of interactions. We see that for 
elongation $\eta=1$ which corresponds to ensemble ${\cal E}_1$ the ground
state is not in the vicinity of any two-pion state and thus it is mainly a $q\bar q$ state, whereas the first two excited
states are close to non-interacting two-pion states, which indicates that they have large two-hadron components.
That is the reason why the multi-hadron operators $\pi\pi_{001}$ and $\pi\pi_{011}$ are required to 
extract these states reliably. For ${\cal E}_1$ and ${\cal E}_2$ these multi-hadron operators are sufficient.
However, for $\eta =2.0$, the second excited state is no longer near the non-interacting pions moving with
back-to-back momentum $\bm p=(0,1,1)$, because the state with back-to-back momentum $\bm p=(0,0,2)$
has lower energy for this elongation. Note that this level crossing is kinematical in nature rather than due to a 
resonance. This is a peculiar feature of our geometry due to the fact that the ordering of levels with different
transverse momenta changes when going from small elongations to large ones. Thus, in order to extract the 
second excited state reliably on ${\cal E}_3$ and ${\cal E}_6$ we need to add the $\pi\pi_{002}$ interpolating
field to our basis. For these ensembles we use a $7\times7$ correlation matrix and extract four energy levels
since the third excited state is very close the the second excited state and below the $4m_{\pi}$ threshold. 
As a result, we will have more data points to fit for the phase shift pattern in next section. The number of energy 
levels we extracted for each ensemble is listed in Table~\ref{Tab:fit_details} in Appendix~\ref{appendix:fitting}.

\section{Resonance Parameters}
\label{sec:rho}

We extract the resonance parameters by fitting the phase-shift data, or equivalently the energy levels,
using two fitting forms: a simple Breit-Wigner form and a model based on Unitarized Chiral Pertubation
Theory (\uchipt). Note that when fitting the phase shift data, the correlation between $E_{\rm cm}$ and $\delta(E_{\rm cm})$ has to be taken into account. The Breit-Wigner form is used in most lattice studies of the $\rho$-resonance since it fits the
phase-shift well. This also offers a straightforward way to compare our results with the ones from other 
studies. The \uchipt\ model provides an alternative parametrization which
also captures well the phase-shift behavior in the $\rho$-resonance region. Its main advantage, and the
reason we use it in our study, is that it can be used to fit the data sets at different quark masses
simultaneously, and it offers a reasonable way to extrapolate our results to the physical point.

The Breit-Wigner parametrization is described in Eq.~\ref{eq:bw-param}. In Fig.~\ref{fig:phase-shift} we
show our phase-shifts and the fitted curves. Note that if we try to fit the entire elastic region, $E_\text{cm}<4 m_\pi$,
the quality of the fit, as indicated by $\chi^2$ per degree of freedom, is not very good. While the curve 
passes close to our points, our energy level determination is very precise and the Breit-Wigner form
is not describing the entire energy range accurately. This is not a serious problem since the Breit-Wigner
form is only expected to describe the data well near the resonance. Ideally, we would restrict the fit
only to a narrow region around the resonance, but the number of data points included in our fit is also
reduced and the fit is poorly constrained. As a compromise we decided to fit the data points that fall
in the range $E_\text{cm}\in[m_\rho-2\Gamma_\rho, m_\rho+2\Gamma_\rho]$. The fit quality is improved and
we will use the results of the narrower fits in the following discussion. For the heavier pion mass the
results in lattice units are
\beq
am_\pi= 0.1934(5),\,am_\rho=0.4878(4),\,g_{\rho\pi\pi}=5.47(11)
\eeq 
and in physical units
\beqs
&m_\pi= 315.4(0.8)(6.3)\MeV\,,\quad m_\rho=795.5(0.7)(16)\MeV\,,\\
&\Gamma_\rho=35.7(1.4)(0.7)\MeV\,,\quad\Gamma'_\rho=124.4(5)(2.5)\MeV\,,
\eeqs
where $\Gamma_\rho$ is the width at the current pion mass and $\Gamma'_\rho$
is the width extrapolated to the physical point. The widths are evaluated using
Eq.~\ref{eq:bw-param} with $\Gamma_\rho=\Gamma_{BW}(m_\rho,m_\pi)$ and 
$\Gamma'_\rho=\Gamma_{BW}(m_\rho^\text{phys},m_\pi^\text{phys})$.
The first error is the
stochastic error and the second one is the systematic error due to the lattice
spacing determination. For the lighter pion mass we have
\beq
am_\pi= 0.1390(5)\,,am_\rho=0.4613(10)\,,g_{\rho\pi\pi}=5.69(12)\,,
\eeq
and in physical units
\beqs
&m_\pi= 225.7(0.8)(4.5)\MeV\,,\quad m_\rho=749.2(1.6)(15)\MeV\,,\\
&\Gamma_\rho=81.7(3.3)(1.6)\MeV\,,\quad\Gamma'_\rho=134.4(5)(2.7)\MeV\,.
\eeqs

We note that the Breit-Wigner fit parameters depend mildly on the range of the fit. The mass 
of the resonance is very well determined, with stochastic errors of the order of few parts
per thousand, and it is insensitive to the fit range. This is because the place where the
phase-shift passes through $\pi/2$ is well constrained by the lattice data. The coupling
$g_{\rho\pi\pi}$ is only constrained at the level of two percent and it is more sensitive
to the fit range, showing a clear drift towards lower values as we narrow the fitting range.

If we are interested in capturing the phase-shift behavior in the entire energy
range available, we could use slight variations of the Breit-Wigner parametrization. Indeed we found that the quality of the fit in the full elastic region is improved when adding 
barrier terms~\cite{VonHippel:1972fg}, especially on the larger pion mass ensemble.
However, such fitting forms change the way the resonance mass and width are
defined making it harder to compare our results directly with other determinations 
and we will not discuss these results here. We include all the relevant data for the
extracted energies and their correlation matrix in Appendix~\ref{appendix:fitting} and 
invite the interested reader to use it to fit any desired parametrization. 

For the Breit-Wigner fit we found that the quality of the fit changes significantly as 
we vary the pion mass within its error bounds. If the Breit-Wigner fit was known to be
the exact description of the phase shift in the elastic region, we could in principle
use the pion mass as a fitting parameter in this fit to further constrain its value.
Since this is not the case, we did not attempt to do this here.

We turn now to the discussion of the fit using the \uchipt\
model. A description is provided in Appendix~\ref{appendix:uchipt}.
An important feature is that this model can be used to fit the phase-shift for both
quark masses {\em simultaneously}. This allows us to extrapolate the results to
the physical point and also to assess the corrections due to the missing strange
quark mass in our calculation. When considering only the $\pi$-$\pi$ channel, 
the model requires as input the pion mass, the pion decay constant and two 
low-energy constants, $\hat l_{1,2}$. The pion mass and decay constants used are the ones in
Table~\ref{Tab:ensembles}. Note that the model can take directly dimensionless
input--$am_\pi$, $af_\pi$ and the energies $aE$---so the systematic errors associated 
with the lattice spacings play no role in the extraction of dimensionless parameters $\hat l_{1,2}$. 
The error-bars that appear in the tables below reflect just the stochastic error. 

\begin{table}[t]
\begin{tabular*}{0.99\columnwidth}{@{\extracolsep{\stretch{1}}}c*{5}{r}@{}}
\toprule 
$m_\pi\,[\MeV]$& $\hat l_1\times 10^3$ & $\hat l_2\times 10^3$ & $m_\rho[\MeV]$ & $\Gamma_\rho[\MeV]$  & $\chi^2/\text{dof}$\\ 
\midrule
315 & 1.5(5) & -3.7(2) & 796(1) & 35(1) & 1 \\
138 &           &            & 704(5) & 110(3)& \\
\cmidrule{1-6}
226 & 2(1) & -3.5(2) & 748(1) & 77(1) & 1.53 \\
138 &          &            & 719(4) & 120(3) &  \\
\cmidrule{1-6}
combined & 2.26(14) & -3.44(3) &  &  & 1.26 \\
138 &          &            & 720(1) & 120.8(8) &  \\
\bottomrule
\end{tabular*}
\caption{\uchipt\ fits in the 
$m_\rho\pm 2\Gamma_\rho$ region and extrapolations to the physical point.
The errors quoted are statistical. The upper two entries show the cases of heavy and light pion mass, 
both individually extrapolated to the physical point. 
The third entry shows the combined fit of both masses and its extrapolation.}
\label{table:uchiptfit1}
\end{table}

In Table~\ref{table:uchiptfit1} we show the results of fitting the \uchipt\ model.
The model is similar to the Breit-Wigner parametrization: it captures the broad 
features of the phase-shift in the elastic region but the quality of the fit is
not good when trying to fit all energy range. We restrict the fit range to 
$m_\rho\pm2\Gamma_\rho$, as we did for Breit-Wigner parametrization. In this
range the quality of the fit is reasonable. The resonance mass is determined
from the center-of-mass energy that corresponds to a $90^\circ$ phase-shift. The
width corresponds to the imaginary value of the resonance pole in the complex
plane. While these parameter definitions are not the same as the ones
determined from the Breit-Wigner fit, the results are consistent as can be seen
from the table.

Fitting each quark mass separately
produces consistent values for $\hat l_{1,2}$ which indicates that the phase-shift
dependence on the quark mass is well captured by this model. Since the model
is consistent for both quark masses we can do a combined fit which allows us to pin
down $\hat l_{1,2}$ with even better precision. As can be seen from the table
the combined fit quality is similar to the individual ones. We will use these
parameters in the subsequent discussion. 

Moreover, we can try to estimate the effects due to the strange quark
using the \uchipt\ model by turning on the coupling to the K$\bar{\rm K}$ channel. We fix the $\pi\pi\to K\bar K$ 
and $K\bar K\to K\bar K$ transitions from a fit to the physical data, while keeping $\hat l_{1,2}$ for the $\pi\pi$ transition at the
values we got from fitting our data. The pion decay constant is adjusted to mach the values in 
Table~\ref{Tab:ensembles}. We report these estimates in Table~\ref{table:uchiptfit2}.
More details about the \uchipt\ fit are included in Appending~\ref{appendix:uchipt}.


\begin{figure}[t]
\includegraphics[width=0.99\columnwidth]{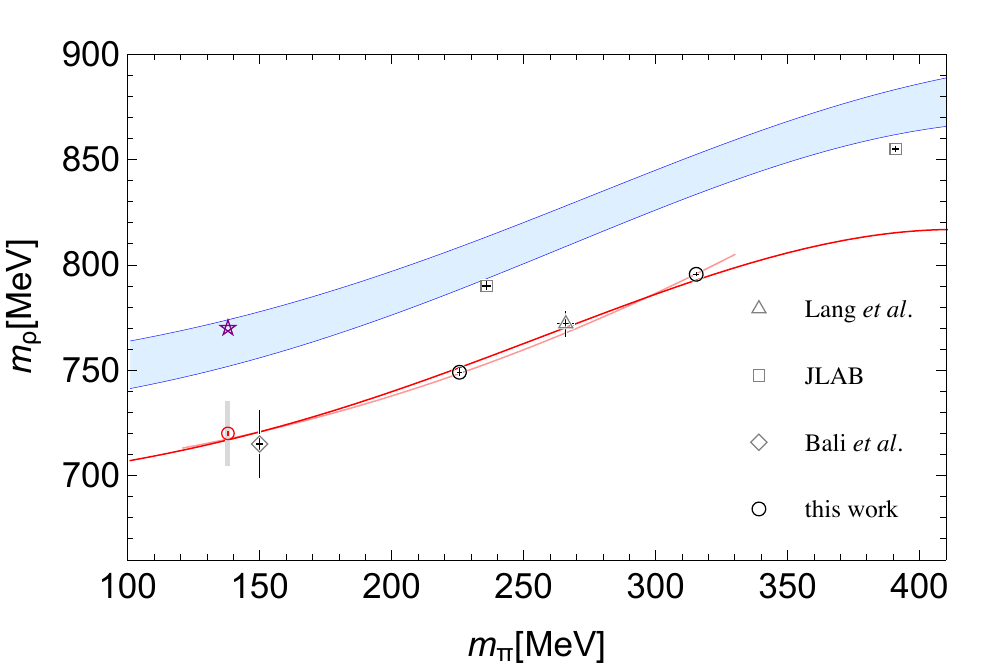}
\caption{Resonance mass extrapolation to the physical point. The red curve corresponds
to an extrapolation based on the \uchipt\ model. The light-red curve corresponds to 
a simple $m_\rho = (m_\rho)_0 + \text{const}\times m_\pi^2$ fit~\cite{Bruns:2004tj}. The blue band
corresponds to an $N_f=2+1$ estimate based on the \uchipt\ model (see text). The other
lattice data-points are taken from Lang {\it et al}~\cite{Lang:2011mn},
JLab group studies~\cite{Dudek:2012xn,Wilson:2015dqa}, and Bali~{\it et al}~\cite{Bali:2015gji}.
The star corresponds to the physical result. The error-bars shown with solid lines are stochastic.
For the extrapolation the gray, thick error-bar indicates the systematic error associated with
the lattice spacing determination.
}
\label{fig:extrapolation}
\end{figure}

In Fig.~\ref{fig:extrapolation} we plot our results for the resonance mass together with 
the \uchipt\ extrapolation, in comparison with results from other lattice groups. 
It is clear that the extrapolation to the physical point in SU(2) is 
significantly below the experimental value, missing it by about $50\MeV$ which is about
8\% of the resonance mass. The stochastic error for the extrapolated result is tiny compared 
with the shift. The systematic error due to the lattice spacing determination is larger, 
but even this cannot account for the discrepancy. The other possible sources of systematic 
errors are finite lattice spacing contributions, finite volume corrections, quark mass 
extrapolation error, and systematics associated with the missing K$\overline{\rm K}$
channel. The lattice artifacts errors are included in our estimate for the systematic
error associated with the lattice spacing determination. To gauge the effect of the lattice
volume corrections we compare our results with the ones from a study by Lang~{\it et al}~\cite{Lang:2011mn}.
This study was carried out on boxes of volume $(2\fm)^3$, whereas our study uses boxes of about
$(3\fm)^3$. We see in Fig.~\ref{fig:extrapolation} that the results agree and we conclude that the finite
volume corrections cannot account for the discrepancy either. The errors associated with
the quark mass extrapolation are also expected to be small: in Fig.~\ref{fig:extrapolation} we
show the results of the extrapolation using a simple polynomial extrapolation which at leading
order depends on $m_\pi^2$~\cite{Bruns:2004tj}\footnote{There is lattice QCD evidence that the 
mass of $\rho$ is well described by a linear dependence on $m_\pi^2$ near the physical quark mass 
below $370\MeV$~\cite{Chen:2015tpa}.}.
 The extrapolation agrees well with the
prediction of \uchipt\ in SU(2). Moreover, a recent calculation by Bali~{\it et al}~\cite{Bali:2015gji}
close to the physical quark mass is also consistent with our extrapolation. 

\begin{table}[b]
\begin{tabular*}{0.99\columnwidth}{@{\extracolsep{\stretch{1}}}c*{5}{r}@{}}
\toprule 
$m_\pi\,[\MeV]$& $m_\rho[\MeV]$ & $\Gamma_\rho[\MeV]$  & $\hat m_\rho[\MeV]$ & $\hat \Gamma_\rho[\MeV]$  &\\ 
\midrule
315 & 795.2(7) & 36.5(2) & 846(0.3)(10) & 54(0.1)(3) \\
226 & 747.6(6) & 77.5(5) & 793(0.4)(10) & 99(0.3)(3) \\
138 & 720(1) & 120.8(8) & 766(0.7)(11) & 150(0.4)(5) \\
\bottomrule
\end{tabular*}
\caption{\uchipt\ results for $N_f=2$, $m_\rho$ and $\Gamma_\rho$, and $N_f=2+1$ estimates,
$\hat m_\rho$ and $\hat \Gamma_\rho$.
The parameters $\hat l_{1,2}$ are taken from the combined fit and the K$\overline{\rm K}$ channel
parameters are taken from fits to experimental data.
The first set of errors quoted are statistical; for $\hat m_\rho$ and $\hat\Gamma_\rho$ we also quote
a set of systematic errors associated with model dependence (see Appendix~\ref{appendix:uchipt} for details).}
\label{table:uchiptfit2}
\end{table}

The likely reason for the discrepancy between the extrapolation and the
physical result is the fact that the strange quark flavor is not included in our calculation.
We note first that the results for \uchipt\  in SU(2) agree very well with the results of in the other $N_f=2$
studies by Lang and Bali.  The results when the strange quark is included are also shown in Fig.~\ref{fig:extrapolation} (blue band indicating
estimated model uncertainties as discussed in Appendix~\ref{appendix:uchipt}). Note that the estimated shift
is surprisingly large and it reduces the discrepancy substantially. The estimated resonance mass curve agrees
quite well with a $N_f=2+1$ lattice calculations reported by the Jlab 
group~\cite{Dudek:2012xn,Wilson:2015dqa}. While these estimates
are likely affected by systematic errors, we feel that they are accurate enough to indicate
that the discrepancy is mostly generated by the absence of the strange quark in our
calculation. 
We note that the
magnitude of the shift in the resonance mass due to the inclusion of the K$\bar{\rm K}$ 
channel is surprisingly large. The present work stresses the importance
of taking into account $\pi\pi-K\bar{K}$ loops, which is the strength of the prediction of the U$\chi$PT model. We will
discuss this point in detail in an upcoming publication~\cite{Hu:2016xx}.


\begin{figure*}
\includegraphics[width=0.49\textwidth]{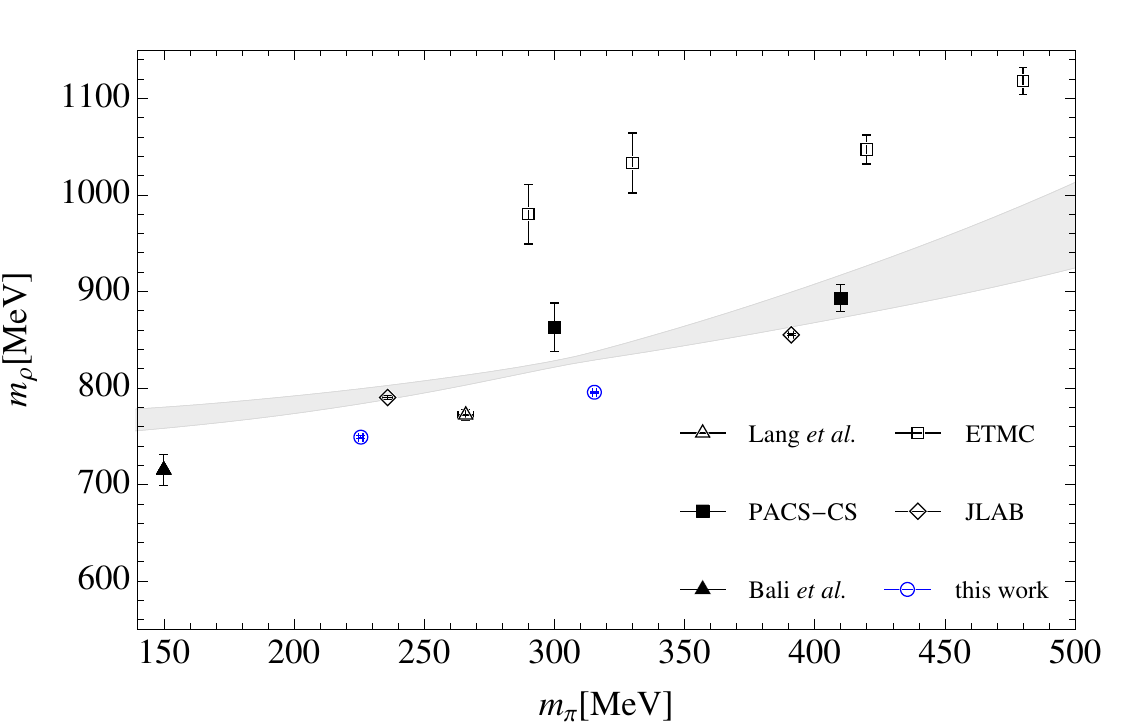}
\includegraphics[width=0.47\textwidth]{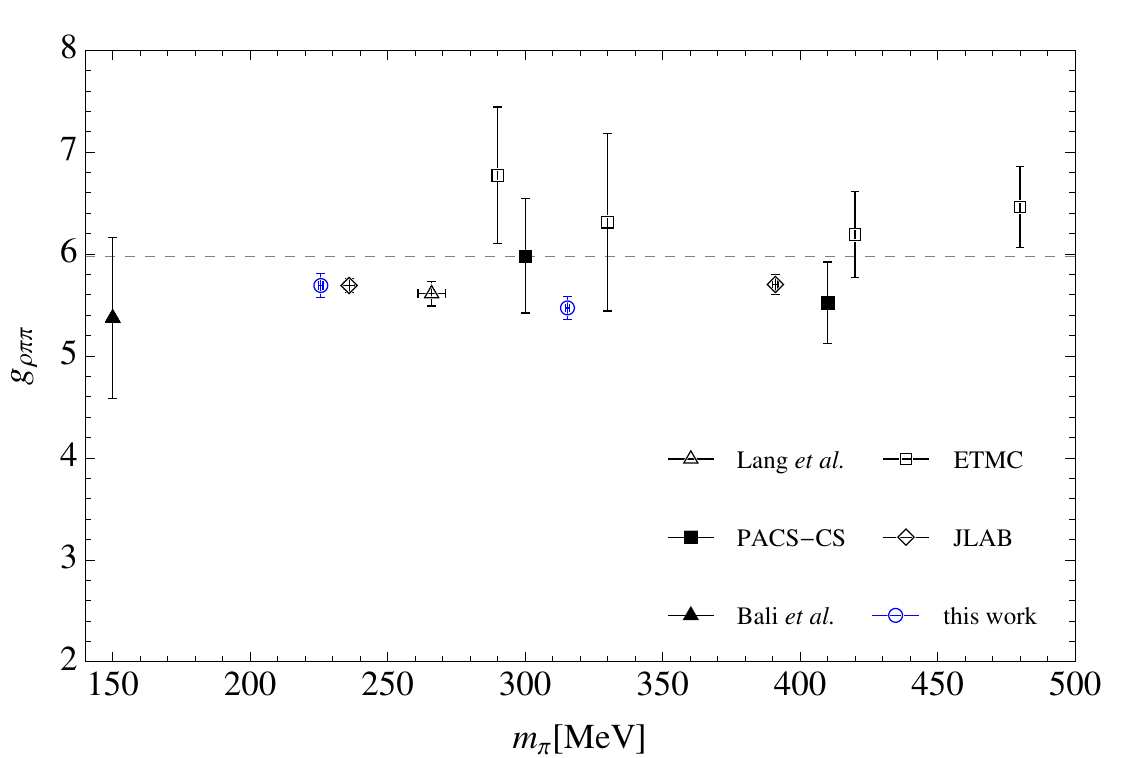}
\caption{Comparison of different lattice calculation for the $\rho$ resonance
mass (left) and width parameter $g_{\rho\pi\pi}$ (right). The errors included
here are only stochastic. The results labeled ETMC are taken from~\cite{Feng:2010es} and
PACS from~\cite{Aoki:2011yj}. The band in the left plot indicates a $N_f=2+1$ expectation from 
\uchipt\ model constrained by some older lattice QCD data and some other physical
input~\cite{Pelaez:2010fj}. The dotted line in the right plot indicates the $g_{\rho\pi\pi}$
that corresponds to the physical rho width~\cite{Agashe:2014kda}.} 
\label{fig:compare}
\end{figure*}

\section{Conclusions}
\label{sec:conclusion}


We presented a high-precision calculation of the phase-shift in the $I=1$, $J=1$
channel for $\pi\pi$ scattering. To scan the resonance region we elongated the
lattice in only one direction, which makes the generation of configuration less
expensive. We used two sets of ensembles, each with three different elongations,
for two different quark masses. To compute the phase-shift we extracted the
energies in the $A_2^-$ channel both for states at rest and for states with
one unit of momentum in the elongated direction. The required two-,
three-, and four-point correlation functions were computed using the LapH method.
Elongated boxes have a different symmetry than cubic ones and 
different L\"uscher formulas are required: for the zero-momentum case
they were worked out in~\cite{Feng:2004ua,Li:2003jn} and in this paper 
we worked out the required one for states boosted along the elongated direction.

The phase-shifts are broadly described by a Breit-Wigner parametrization, as
expected. However, our calculation is precise enough to show that more
sophisticated models are required to describe the variation of the 
phase-shift in the entire elastic region. It is hoped that our results can
be used to validate these models and to constrain their parameters.

The resonance mass and $g_{\rho\pi\pi}$ coupling are extracted from 
fitting a Breit-Wigner parametrization in the energy region $m_\rho\pm2\Gamma_\rho$.
In Fig.~\ref{fig:compare} we compare our results with other lattice determinations.
Other than the ETMC study, the lattice data seems to be split in two groups:
$N_f=2+1$ results (PACS~\cite{Aoki:2011yj} and Jlab~\cite{Dudek:2012xn,Wilson:2015dqa})
which are in agreement with $N_f=2+1$ expectations from \uchipt~\cite{Pelaez:2010fj},
and $N_f=2$ lattice data (Lang~{\it et al}~\cite{Lang:2011mn}, Bali~{\it et al}~\cite{Bali:2015gji},
and this work) that agrees with a $N_f=2$ \uchipt\ model fit to our data.

For the resonance mass, we performed an extrapolation to the physical
mass using a \uchipt\ model, which we found can describe well the phase-shift
data at both quark masses using the same parameters. The extrapolation results 
are consistent with extrapolations based on other models~\cite{Bruns:2004tj} and
other $N_f=2$ lattice calculations, as discussed before. The extrapolated results differ significantly
from the physical one and we argue that this is due to the absence of the strange
quark in our calculation.

For the quark masses used in this study, we did not find evidence of significant finite volume
effects. However, as we lower the pion mass, larger volumes would most likely be required.
The original LapH method might turn out to be too expensive to apply directly, but a
stochastic variant was already developed and showed to work well~\cite{Bulava:2016mks}.
Note that we did not include these data points in Fig.~\ref{fig:compare} because this
study is using the same ensemble as the Jlab group study~\cite{Wilson:2015dqa} and their
results are compatible with the ones computed by the Jlab group, albeit with slightly larger
errorbars.

Turning to the future, as far as phase-shifts in the $\rho$ resonance channel are
concerned, we seem to have moved beyond proof-of-principle calculations and toward
precision determinations. We anticipate that in the near future precise calculations
at the physical point might be possible that will give us access directly to phase-shifts
to be compared to values extracted from physical data. For example, it would be interesting
to see whether phaseshifts close to threshold match chiral perturbation theory expectations,
which at leading order are controlled solely by $m_\pi$ and $f_\pi$. 
The phaseshifts calculated in this study do not agree well with this prediction even for the
data points closest to the threshold. Note that even the experimental determined phases also
do not agree with the lowest order chiral perturbation predictions. 

Other channels are also of
interest, for example the $I=0$ channel where the broad sigma resonance is expected to appear,
$\pi$-K scattering in the $K^*$ channels, baryon-meson scattering, etc. We plan to investigate 
some of these channels in the near future.

\acknowledgements

D.G. and A.A. are supported in part by the National Science Foundation CAREER grant PHY-1151648. 
M.D. and R.M. are supported by the National Science Foundation (CAREER grant PHY-1452055, 
PIF grant No. 1415459) and by GWU (startup grant). The authors thank P. Bedaque, G. Bali, R. Brice\~no, D. Mohler, and S. Prelovsek
for discussions and correspondence related to this project. A.A. gratefully acknowledges 
the hospitality of the Physics Department at the University of Maryland where part of this work was carried out.
The computations were carried out on GWU Colonial One computer cluster and GWU IMPACT collaboration clusters;
we are grateful for their support.

\appendix
\section{Zeta function}
\label{appendix:zeta_func}

To compute the phase shift in Eq.~\ref{eq:phaseshift_formula} we need to numerically evaluate the zeta function. 
For the zero momentum case the relevant formulas for elongated boxes were derived in~\cite{Li:2003jn,Feng:2004ua}. 
In the following we will show how to extend this to the non-zero momentum states, when the boost is parallel 
with the elongated direction of the box. We discuss first the evaluation of the zeta function for cubic boxes and
then extend it to accommodate elongated boxes. For a boosted state with momentum $\bm P = \bm d 2\pi/L$, the zeta 
function in a cubic box is
\beq
\mathcal{Z}_{lm}^{\bm d}(s;q,\gamma) = \sum_{\bm n\in P_{\bm d}(\gamma,\eta)} \frac{\mathcal{Y}_{lm}(\bm n)}{(\bm n^2-q^2)^s},
\eeq
with 
\beq
P_d(\gamma)=\left\{\bm n\in \reals^3 \Big| \bm n=\hat{\gamma}^{-1}\left({\bm m}+\frac{\bm d}{2}\right),\bm m\in {\mathds Z}^3\right\}.
\eeq
The series above is only convergent when $\Re s>l+3/2$ but the zeta function needs to be evaluated at $s=1$. 
The function defined by the series above can be analytically continued in the region $\Re s>1/4$.
The analytic continuation is done following L\"{u}scher~\cite{Luscher:1990ux} and Rummukainen~\cite{Rummukainen:1995vs} 
using the heat kernel expansion
\beqs
\mathcal{K}(t,\bm x) &= \frac{1}{(2\pi)^3}\sum_{\bm n\in P_{\bm d}(\gamma)} e^{i\bm n\cdot \bm x -t \bm n^2}\\
&= \frac{\gamma}{(4\pi t)^{\frac{3}{2}}} \sum_{\bm n\in P_{\bm 0}(1/\gamma)} 
e^{i 2\pi(\hat{\gamma}^{-1} \bm n)\cdot \bm d/2} e^{-\frac{1}{4t} (\bm x +2\pi\bm n)^2} \,.
\eeqs
This relation is obtained using Poisson's summation formula 
\beq
\sum_{n\in Z^3}f(\bm n)=\sum_{k\in Z^3} \int^{\infty}_{-\infty}f(\bm x)e^{i 2\pi\bm k\cdot\bm x} d^3 x.
\eeq
The spherical projected kernel $\mathcal{K}_{lm}$ is defined as 
$\mathcal{K}_{lm}= \mathcal{Y}_{lm} (-i\nabla)\mathcal{K}$ which can be written as 
\beqs
&\mathcal{K}_{lm}^{\bm d}(t,\bm x)=(2\pi)^{-3/2}\sum_{\bm n\in P_{\bm d}} \mathcal{Y}_{lm}(\bm n) e^{(i \bm n\cdot \bm x -t \bm n^2)}\\
&=\frac{\gamma}{(4\pi t)^{3/2}}\frac{i^l}{(2t)^l}\sum_{\bm n\in Z^3}(-1)^{\bm d\cdot\bm n} 
\mathcal{Y}_{lm}(\bm x+2\pi\hat{\gamma} \bm n) e^{-\frac{1}{4t}(\bm x+2\pi \hat{\gamma} \bm n)^2} \,.
\eeqs
Using the truncated kernels $\mathcal{K}_{lm}^\Lambda= \mathcal{Y}_{lm} (-i\nabla)\mathcal{K}^\Lambda$ with
\beq
\mathcal{K}^{\Lambda}(t,\bm x)\equiv\frac{1}{(2\pi)^3}\sum_{\bm n\in P_{\bm d}(\gamma),|\bm n|\le\Lambda} e^{i\bm n\cdot \bm x-t\bm n^2}\,,
\eeq
we define the zeta function by separating the series terms in two groups, a finite set close to the
origin that remains in the original form, and the rest that will be evaluated via an kernel integral:
\beqs
&\mathcal{Z}_{lm}^{\bm d}(s;q,\gamma) \\
&=\sum_{\substack{\bm n\in P_{\bm d}(\gamma)\\ |\bm n|<\Lambda}} \frac{\mathcal{Y}_{lm}(\bm n)}{(\bm n^2-q^2)^s}+\frac{(2\pi)^3}{\Gamma(s)}\left\{
\frac{\delta_{l0}\delta_{m0} \gamma}{(4\pi)^2(s-3/2)}\right. \\
&+\int^1_0dtt^{s-1}\left[e^{tq^2}\mathcal{K}_{lm}^{\Lambda}(t,\bm 0)-\frac{\delta_{l0}\delta_{m0}\gamma}{(4\pi)^2t^{3/2}}\right] \\
&+\left.\int^{\infty}_1 e^{tq^2}\mathcal{K}_{lm}^{\Lambda}(t,\bm 0)\right\} \,.
\eeqs
For the integral on the $t\in [1,\infty]$ range, the heat kernel expansion in 
terms of $\exp(i \bm n\cdot\bm x-t \bm n^2)$ is used, and on the $t\in[0,1]$ range, 
the expansion in terms of $\exp[-(\bm x+2\pi\bm n)^2/4t]$ is used. In both cases the series 
converges the slowest around $t=1$: for the $t\in [1,\infty]$ range large $\bm n$ terms contribute 
$\exp(-t(\bm n^2-q^2))$ and on the $t\in [0,1]$ range they contribute $\exp[tq^2-(2\pi \bm n)^2/4t]$. 
It is clear that this series converges quickly for large $\bm n$ if we choose $\Lambda> \Re q^2$. 
For the $A_2^-$ irrep we need to evaluate $\mathcal{Z}^{\bm d}_{00}$ and $\mathcal{Z}^{\bm d}_{20}$. 
For $\mathcal{Z}_{00}^{\bm d}$, we have 
\beqs
&\mathcal{Z}^{\bm d}_{00}(s;q,\gamma)\\
&=\sum_{\substack{\bm n\in P_{\bm d}(\gamma)\\|\bm n|\le\Lambda}} \frac{\mathcal{Y}_{00}(\bm n)}{(\bm n^2-q^2)^s}+\frac{\pi\gamma}{2\Gamma(s)(s-3/2)}\\
&+\gamma\sum_{\bm n\in P_{\bm 0}(1/\gamma)} e^{i 2\pi(\hat{\gamma}^{-1}\bm n)\cdot \bm d/2} \mathcal{Y}_{00}(\bm n)I_{00}(s,q,|\bm n|)\\
&-\sum_{\substack{\bm n\in P_{\bm d}(\gamma)\\|\bm n|\le \Lambda}} \mathcal{Y}_{00}(\bm n)\underline{J}_{00} (s,q,|\bm n|)\\
&+\sum_{\substack{\bm n\in P_{\bm d}(\gamma)\\|\bm n|>\Lambda}} \mathcal{Y}_{00}(\bm n)\bar{J}_{00}(s,q,\gamma,|\bm n|) \,,
\eeqs
where $\mathcal{Y}_{00}(\bm n)=\frac{1}{\sqrt{4\pi}}$ and the notation used is:
\beqs
I_{00}(s,q,|\bm n|)&=\frac{(2\pi)^3}{\Gamma(s)} \int_0^1 dt t^{s-1} \frac{(e^{tq^2-(2\pi\bm n)^2/4t} - \delta_{|\bm n|0})}{(4\pi t)^{3/2}}\,,\\
\underline{J}_{00}(s,q,|\bm n|)&=\frac{1}{\Gamma(s)}\int^1_0 dt t^{s-1}e^{tq^2}e^{-t\bm n^2}\,,\\
\bar{J}_{00}(s,q,|\bm n|) &=\frac{1}{\Gamma(s)} \int^{\infty}_1 dt t^{s-1} e^{tq^2} e^{-tn^2}\\
&=\frac{1}{(\bm n^2-q^2)^s} -\underline{J}_{00}(s,q,|\bm n|)\,.
\eeqs
These functions can be expressed in terms of Euler gamma function and 
exponential integral function 
\beqs
\Gamma(s,z) = \int_z^{\infty} t^{s-1} e^{-t} dt \,,\quad
E_n(z)      =\int_1^{\infty} \frac{e^{-zt}}{t^n}dt \,.
\eeqs
We have
\beqs
\underline{J}_{00}(s,q,|\bm n|) &= \frac{1}{(\bm n^2-q^2)^s}\left(1-\frac{\Gamma(s,\bm n^2-q^2)}{\Gamma(s,0)} \right) \,,\\
\bar{J}_{00}(s,q,|\bm n|) &=E_{1-s}(\bm n^2-q^2) \ \text{for} \ \mathbf n^2>q^2 \,.
\eeqs
Therefore, the zeta function can be simplified to 
\beqs
\mathcal{Z}_{00}^{\bm d}(s;q,\gamma)&= \frac{\pi \gamma}{2\Gamma(s)(s-3/2)}+\sum_{\bm n\in P_{\bm n}(\gamma)} \mathcal{Y}_{00}(\bm n)\bar{J}_{00}(s,q,|\bm n|)\\
&+ \gamma\sum_{\bm n\in P_{\bm 0}(1/\gamma)}e^{i2\pi(\hat{\gamma}^{-1}\bm n)\cdot \bm d/2}\mathcal{Y}_{00}(\bm n)I_{00}(s,q,|\bm n|),\\
\mathcal{Z}_{20}^{\bm d}(s;q,\gamma)&=\gamma\sum_{\bm n\in P_{\bm 0}(1/\gamma)}e^{i2\pi(\hat{\gamma}^{-1} \bm n)\cdot \bm d/2} \mathcal{Y}_{20}(\bm n)I_{20}(s,q,|\bm n|)\\
&+\sum_{\bm n\in P_{\bm d}(r)} \mathcal{Y}_{20}(\bm n)\bar{J}_{00}(s,q,|\bm n|) \,.
\eeqs
For the case of interest, $s=1$, the $J_{00}$ and $I_{00}$ integrals are
\beqs
\bar{J}_{00}(1,q,|\bm n|)&= \frac{e^{-(\bm n^2-q^2)}}{\bm n^2-q^2},\\
I_{00}(1,q,0)&=2\pi^{3/2}[1-e^{q^2}-\sqrt{\pi}iq \erf(iq)]  \,,\\
I_{00}(1,q,|\bm n|\not=0)&=\frac{\pi}{|\bm n|}Re[e^{2\pi i q|\bm n|}(1-\erf(iq+\pi|\bm n|))] \,.
\eeqs
Similarly, for $s=1$, the $I_{20}$ integral can simplify to
\beqs
&I_{20}(1,q,|\bm n|)=-\frac{(3+2\bm n^2)e^{q^2-(\pi \bm n)^2}}{2|\bm n|^4\sqrt{\pi}}\\
&+\Re\frac{e^{2\pi i q |\bm n|}[-3+2\pi q|\bm n|(3i + 2\pi q |\bm n|)][1-\text{erf}(iq+\pi|\bm n|)]}{4|\bm n|^5\pi}
\eeqs
where the error function is defined as 
\beq
\erf(z) =\frac{2}{\sqrt{\pi}} \int_0^z e^{-t^2}dt \,.
\eeq
All the relations above work for $\bm n^2 -q^2\neq 0$. The series are divergent at the points 
where $\bm n^2=q^2$. To avoid this trivial divergence these points are removed from the summation, that is 
\beq
\mathcal{Z}_{lm}^{\bm d}(s;q,\gamma)=\sum_{\substack{\bm n\in P_{\bm d}(\gamma)\\
\bm n^2\neq q^2}} \frac{\mathcal{Y}_{lm}(\bm n)}{(\bm n^2-q^2)^s}\,.
\eeq
This basically amounts to replacing $1/(\bm n^2-q^2)$ with $0$ when $\bm n^2=q^2$. 
In the simplified expressions above this is equivalent to setting 
\beq
\bar{J}_{lm}(s,q,|\bm n|)=-\frac{1}{s\Gamma(s)}
\eeq
when $q^2=|\bm n|^2$. This is because the convergent counterpart 
$\underline{J}_{lm}(s,q,|\bm n|)=1/s\Gamma(s)$ for $s>0$ when $q^2=\bm n^2$ and the sum of 
$\bar{J}_{lm}$ and $\underline{J}_{lm}$ is $1/(\bm n^2-q^2)$ which is replaced with $0$. 

For the elongated box case the zeta functions also depend on the elongation factor $\eta$. 
The only difference is the domain of the summation which becomes 
\beq
P_{\bm d}(\gamma,\eta)= \left\{\bm n\in \reals^3\Big|\bm n = \hat{\gamma}^{-1}\hat{\eta}^{-1}(\bm m+\frac{\bm d}{2}),\bm m\in {\mathds Z}^3\right\}\,,
\eeq
where $\hat{\gamma}$ and $\hat{\eta}$ are defined in Eq.~\ref{eq:gamma} and Eq.~\ref{eq:eta}. 
In sum, they can be calculated as 
\beqs
\mathcal{Z}_{00}^{\bm d}(s;q,\gamma,\eta) &= \mathcal{Z}_{00}^{\bm d}(s;q,\gamma\eta)\,,\\
\mathcal{Z}_{20}^{\bm d}(s;q,\gamma,\eta) &= \mathcal{Z}_{20}^{\bm d}(s;q,\gamma\eta)\,.
\eeqs

\section{Unitarized chiral perturbation theory model}
\label{appendix:uchipt}

Chiral Perturbation Theory ($\chi$PT) is successful in describing the meson-meson interaction at low energies~\cite{gasser1,gasser2}. 
However, the convergence of the amplitude expansion in powers of the meson momenta becomes slow when 
the energy increases. Moreover, the perturbative expansion fails in the vicinity of resonances, such as 
$\sigma$ or $\rho$ mesons. To describe the resonant phase shifts and inelasticities extracted from meson-meson scattering, one needs to extend the theory to higher energies. 
Unitarized Chiral Perturbation Theory (\uchipt) is a nonperturbative method which combines constraints from chiral symmetry and its breaking and 
(coupled-channel) unitarity. The method of Ref.~\cite{Oller:1998hw} uses the ${\cal O}(p^2)$ and ${\cal O}(p^4)$ chiral Lagrangians 
together with a coupled-channel scattering equation  which implements unitarity, and is able to describe 
the meson-meson interaction up to about $1.2\GeV$. The resulting amplitudes show 
poles in the complex plane that can be associated with the known scalar and vector resonances. In the context of the 
Inverse Amplitud method~\cite{Truong1,Dobado1,Dobado2,Dobado3,Nebreda}, 
the two-meson scattering equation reads~\cite{Oller:1998hw}
\begin{equation}
T=[I-VG]^{-1}V
\label{eq:tmat2}
\end{equation}
where
\begin{equation}
V=V_2[V_2-V_4]^{-1}V_2\,.
\label{eq:viam}
\end{equation}
In Eq. \ref{eq:tmat2}, $G$ is a diagonal matrix whose elements are the two-meson loop functions, 
evaluated in our case in dimensional regularization in contrast to the cut-off-scheme used in the original model of Ref.~\cite{Oller:1998hw}:
\begin{eqnarray}
G^{DR}_{ii}(E)
&&= i \,  \int \frac{d^4 q}{(2 \pi)^4} \,
\frac{1}{q^2 - m_{1}^2 + i \epsilon} \, \frac{1}{(P-q)^2 - m_{2}^2 + i
\epsilon}\nonumber\\
 &&= \frac{1}{16 \pi^2} \left\{ a(\mu) + \ln
\frac{m_{1}^2}{\mu^2} \right.\left.+\frac{m_{2}^2-m_{1}^2 + E^{2}}{2E^{2}} \ln \frac{m_{2}^2}{m_{1}^2} 
\right. \nonumber \\ &&  +
\frac{p_i}{E}
\left[
\ln(\hspace*{0.2cm}E^{2}-(m_{1}^2-m_{2}^2)+2 p_i E)
\right. 
 \nonumber\\&&+ \ln(\hspace*{0.2cm}E^{2}+(m_{1}^2-m_{2}^2)+2 p_i E) \nonumber\\
&&-\ln(-E^{2}+(m_{1}^2-m_{2}^2)+2p_i E)  \nonumber\\&&\left. \left.- \ln(-E^{2}-(m_{1}^2-m_{2}^2)+2 p_i E) \right]
\right\} ,
\label{eq:gdimre}
\end{eqnarray}
 where $p_i=\frac{\sqrt{(E^2-(m_1+m_2)^2)(E^2-(m_1-m_2)^2)}}{2 E}$ for the channel $i$, $E$ is the center-of-mass energy, and $m_{1,2}$ refers to the masses of the mesons $1,2$ in the $i$ channel. Throughout this study we use $\mu=1$~GeV and a natural value of the subtraction constant $\alpha(\mu)=-1.28$.
 

For the case of the $\pi-\pi$ interaction with $(I=1;L=1)$, the kernel of Eq.~\ref{eq:tmat2}, 
$V(\pi\pi)$, can be expressed as~\cite{Oller:1998hw}
\begin{equation}
 V(\pi\pi)=\frac{-2\,p^2}{3(f_\pi^2-8\,\hat{l}_1 m_\pi^2+4\, \hat{l}_2 E^2)}\ ,\label{eq:vpipi}
\end{equation}
where specific combinations of LECs have been introduced, $\hat{l}_1\equiv 2\,L_4+L_5$ and $\hat{l}_2\equiv2\,L_1-L_2+L_3$. Note that these are not identical to the SU(2) CHPT low-energy constants. The one-channel reduction given by Eq.~\ref{eq:vpipi}, which contains the lowest- and next-to-leading order contributions,
constitutes the fit model for the $N_f=2$ lattice data of this study.

 

\subsection*{Coupled channel case ($\pi\pi-K\bar K$) }

In this section we describe the meson-meson interaction in terms of the partial-wave 
decomposition of the amplitude and apply it to the case of the 
$\pi\pi-K\bar{K}$ system  with quantum numbers $(L=1;I=1)$.
The partial wave decomposition of the scattering amplitude of two spinless mesons with definite 
isospin $I$ can be written as 
\begin{equation}
T_I=\sum_J(2 J+1)T_{IJ}P_J(\cos\theta)\,.
\end{equation}
where
\begin{equation}
T_{IJ}=\frac{1}{2}\int^{1}_{-1} P_J(\cos\,\theta)T_I(\theta)\,\,{\rm d}\cos\theta\,.
\end{equation}
In the case of two coupled channels, $T_{IJ}$ is a $2\times 2$ matrix whose elements $(T_{IJ})_{ij}$ are
related to $S$ matrix elements through the equations (omitting the $I,J$ labels from here on)
\beqs
&(T)_{11}=-\frac{8\pi E}{2ip_1}[(S)_{11}-1]\ ,\quad (T)_{22}=-\frac{8\pi E}{2ip_2}[(S)_{22}-1]\,,
\\
&(T)_{12}=(T)_{21}=-\frac{8\pi E}{2i\sqrt{p_1p_2}}(S)_{12}\,,
\label{eq:sma2}
\eeqs
with $p_1$, $p_2$ the center-of-mas momenta of the mesons in channel $1$ ($\pi\pi$) or $2$ ($K\bar K$) respectively, that is
$p_i=\sqrt{(E/2)^2-m_i^2}$. The $S$-matrix can be parametrized as
\begin{equation}
S=\pmat{\eta e^{2i\delta_1}& i(1-\eta^2)^{1/2}e^{i(\delta_1+\delta_2)}\\
i(1-\eta^2)^{1/2}e^{i(\delta_1+\delta_2)}& \eta e^{2i\delta_2}}\,.
\label{eq:sma}
\end{equation}
The interaction in the $\pi\pi-K\bar{K}$ system, is evaluated from the ${\cal O}(p^2)$ and ${\cal O}(p^4)$ 
Lagrangians of the $\chi$PT expansion~\cite{gasser1,gasser2}. The potentials, $V_2$ and $V_4$, 
projected in $I=1$ and $L=1$ are~\cite{Oller:1998hw}
\beq
V_{2}(E)=-
\pmat{\frac{2p^2_\pi}{3f_\pi^2} & \frac{\sqrt{2}p_K p_\pi}{3 f_Kf_\pi}\\
\frac{\sqrt{2}p_K p_\pi}{3 f_Kf_\pi} & \frac{p_K^2}{3 f_K^2}}
\label{eq:v2c}
\eeq
and 
\beqs
&V_{4}(E)=-1\times\\
&\pmat{
\frac{8p_\pi^2(2\hat{l}_1 m_\pi^2-\hat{l}_2 E^2)}{3f_\pi^4} & \frac{8p_\pi p_K(L_5(m_K^2+m_\pi^2)-L_3 E^2)}{3\sqrt{2}f_\pi^2 f_K^2} \\
\frac{8p_\pi p_K(L_5(m_K^2+m_\pi^2)-L_3 E^2)}{3\sqrt{2}f_\pi^2 f_K^2} & \frac{4p^2_K (10\hat{l}_1 m_K^2+3(L_3-2\hat{l}_2)E^2)}{9f_K^4}}
\,.
\label{eq:v4c}
\eeqs
The two-channel $T$-matrix is evaluated by means of Eq.~\ref{eq:tmat2}. 
Note that the channel transitions in Eqs.~\ref{eq:v2c} and \ref{eq:v4c} depend on four low energy constants, 
$\hat{l}_1$, $\hat{l}_2$, $L_3$ and $L_5$.

\subsection*{Meson-meson scattering in the finite volume and U$\chi$PT model}

In Refs.~\cite{Doring:2011vk,Doring:2011ip,Doring:2009bi}, a formalism has been developed that is equivalent to the L\"uscher framework up to exponentially suppressed corrections. The formalism is summarized
in this section. Given the two-meson-interaction potential, as the $V$
with the ${\cal O}(p^2)$ and ${\cal O}(p^4)$ terms in the $\chi$PT expansion, that is Eqs.~\ref{eq:viam}, 
\ref{eq:v2c} and \ref{eq:v4c}, the scattering amplitude in the finite volume can be written as, 
\begin{equation}
\tilde{T}=[I-V\tilde{G}]^{-1}V\,,
\label{eq:tmatf}
\end{equation}
or $\tilde{T}=[V^{-1}-\tilde{G}]^{-1}$, similarly to Eq.~\ref{eq:tmat2} in the infinite-volume limit. 
In the case of boxes with asymmetry $\eta$ in the $z$ direction, 
$\tilde{G}$ can be evaluated as,
\begin{equation}
\tilde{G}(E)=\frac{1}{\eta L^3} \sum_{\bm q} I(E, \bm q)\,,
\label{eq:gbox}
\end{equation}
where the channel index has been omitted. Here,
\begin{equation}
I(E, {\bf q})= \frac{\omega_1(\bm q) + \omega_2(\bm q)}
{2 \omega_1(\bm q) \omega_2(\bm q)}\frac{1}{E^2-(\omega_1(\bm q)^2 + \omega_2(\bm q)^2)}\,,
\label{eq:iq}
\end{equation}
where $\bm q=\frac{2\pi}{L}(n_x,n_y,n_z/\eta)$.
The sum over the momenta is cut off at $q_\text{max}$. The formalism can also be made independent of 
$q_\text{max}$ and related to the subtraction constant in the dimensional-regularization method, 
$\alpha$ (as in the continuum limit), see Ref~\cite{MartinezTorres:2011pr}, 
\beqs
&\tilde{G}=G^{DR}+\\
&\lim_{q_\text{max} \to \infty} 
\left( \frac{1}{\eta L^3}\sum_{q<q_\text{max}} I(E,\bm q) - \int_{q<q_\text{max}} 
\frac{d^3 q}{(2\pi)^3} I(E,\bm q) \right) \\
&\equiv G^{DR} + \lim_{q_\text{max} \rightarrow \infty} \delta G\,,
\label{eq:gtil}
\eeqs
where $G^{DR}$ stands for the two-meson loop function given in Eq.~\ref{eq:gdimre}.
For energies $E_i$ which correspond to poles of $\tilde{T}$,
i.e., the energy eigenvalues in the finite volume, we can obtain the $T$ matrix in the infinite volume,
\begin{equation}
T=[\tilde{G}(E)-G(E)]^{-1}
\label{eq:difg}
\end{equation}
which is independent of the renormalization of the individually divergent expressions.

In the general multi-channel case, the energy spectrum in a box, predicted by U$\chi$PT, is found as solution of the equation 
\begin{equation}
\mathrm{Det}[V^{-1}(E)-\tilde{G}(E)]=0 \ .
\end{equation}


As has been shown in Ref.~\cite{Doring:2011vk}, the formalism of Refs.~\cite{Doring:2011vk,Doring:2011ip,Doring:2009bi} is 
equivalent to the L\"uscher approach up to contributions which are exponentially suppressed with the volume. 
In what follows, we refer to Ref.~\cite{Doring:2012eu} for the generalization of the formalism to moving 
frames. The formalism of Ref.~\cite{Doring:2012eu} is generalized to include partial wave mixing and coupled channels, but in the current study the $F$ wave is neglected. 

For an equal-mass system interacting in $p$-wave and moving 
with $\bm P=\frac{2\pi}{\eta L}(0,0,1)$ in the direction of the elongation of the box, we find the following relations,
\begin{eqnarray}
 &&A_2^-:\qquad-1+V(\pi\pi)\tilde{G}_{10,10}=0\label{eq:sym1}\\
 &&E^-:\qquad-1+V(\pi\pi)\tilde{G}_{11,11}=0\ ,\label{eq:sym2}
\end{eqnarray}
with
$\tilde{G}_{lm,l'm'}$ given in Ref.~\cite{Doring:2012eu} but modified as in Eqs.~\ref{eq:iq} and \ref{eq:gtil} by the elongation factor $\eta$. Above, $V(\pi\pi)$ is from
Eq.~\ref{eq:vpipi}. The above relations are used to fit $\hat l_1,\,\hat l_2$ directly to the energy levels in the finite volume.

We have also checked that the numerical results for the phase shifts derived from Eq.~\ref{eq:sym1} are very similar to those in Appendix~\ref{appendix:zeta_func} when the argument of the integrand $I(E,\bm q\,)$ from Eq.~\ref{eq:iq} is replaced as described in Ref.~\cite{Doring:2011vk}, $I(E,\bm q\,)\to (2E)^{-1}(p^2-\bm q^2)^{-1}$, to remove exponentially suppresed contributions and ensure comparability with the L\"uscher formalism. See also Eq.~18 of Ref.~\cite{Doring:2012eu} for the replacement in case of moving frames. In any case, these exponentially suppressed contributions are small in the present case.

\subsection*{\uchipt\ fit results}

\begin{figure}[b]
\begin{center}
\includegraphics[width=\columnwidth]{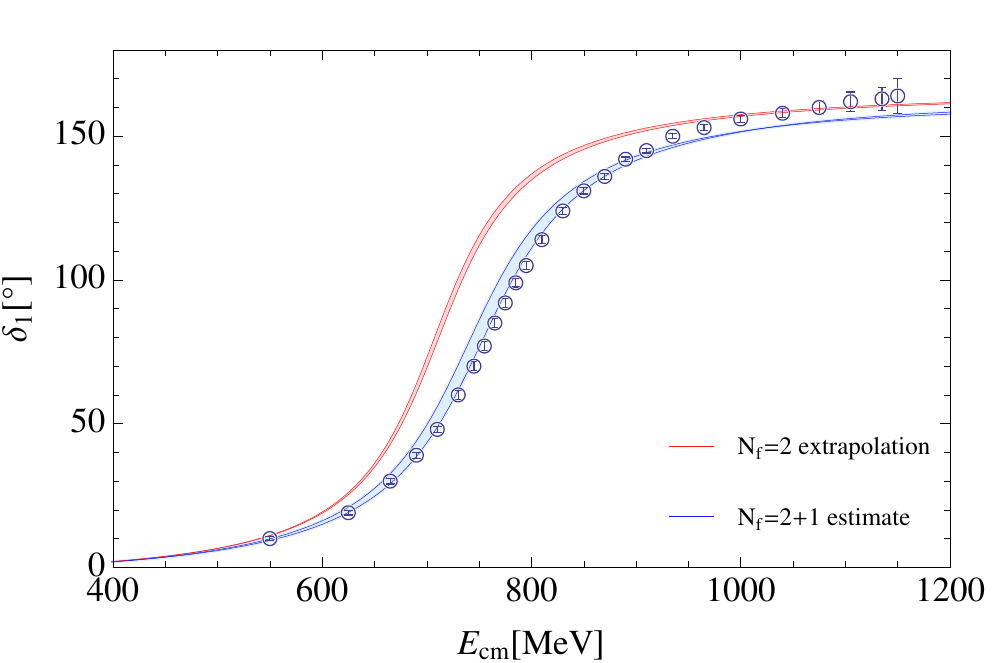}
\end{center}
\caption{Chiral extrapolation of the phase shift to
the physical mass (red band), obtained from the simultaneous fit to lattice eigenvalues at both considered pion masses. Only statistical uncertainties are indicated. The blue band
shows the estimated phase shift when including also the $K\bar{K}$ channel in the two variants mentioned in the text. To keep the figure simple, statistical uncertainties are not indicated for these cases. They are of the same size as the red band. 
Open circles indicate phase shifts extracted from experiment~\cite{Protopopescu:1973sh}.}
\label{fig:fitfincb}
\end{figure}

\begin{figure*}[t]
\begin{center}
\begin{tabular}{ll}
\includegraphics[width=0.475\textwidth]{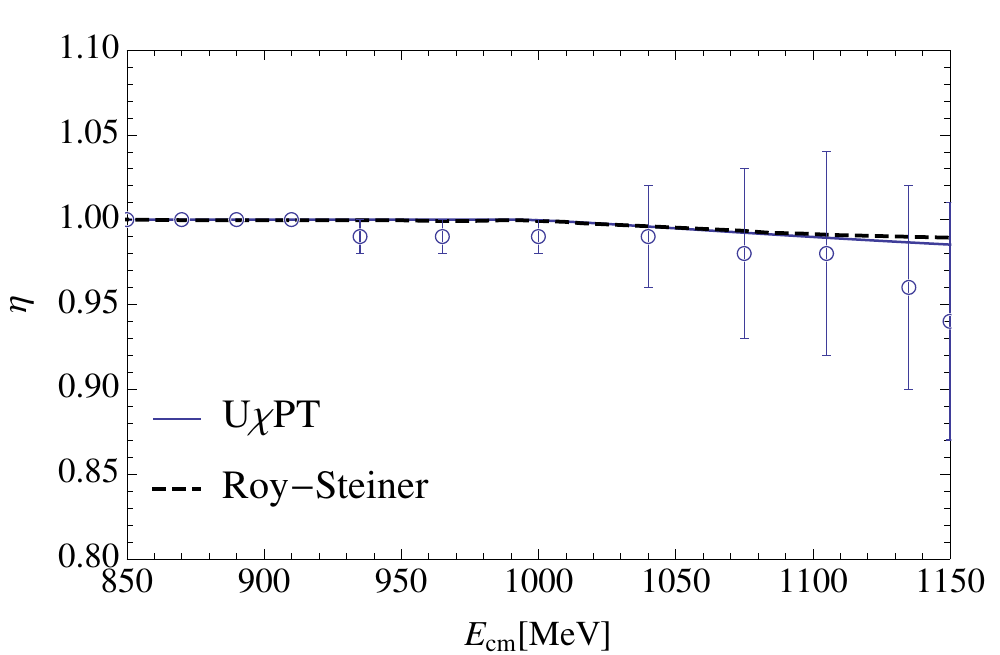}\kern5mm
\includegraphics[width=0.475\textwidth]{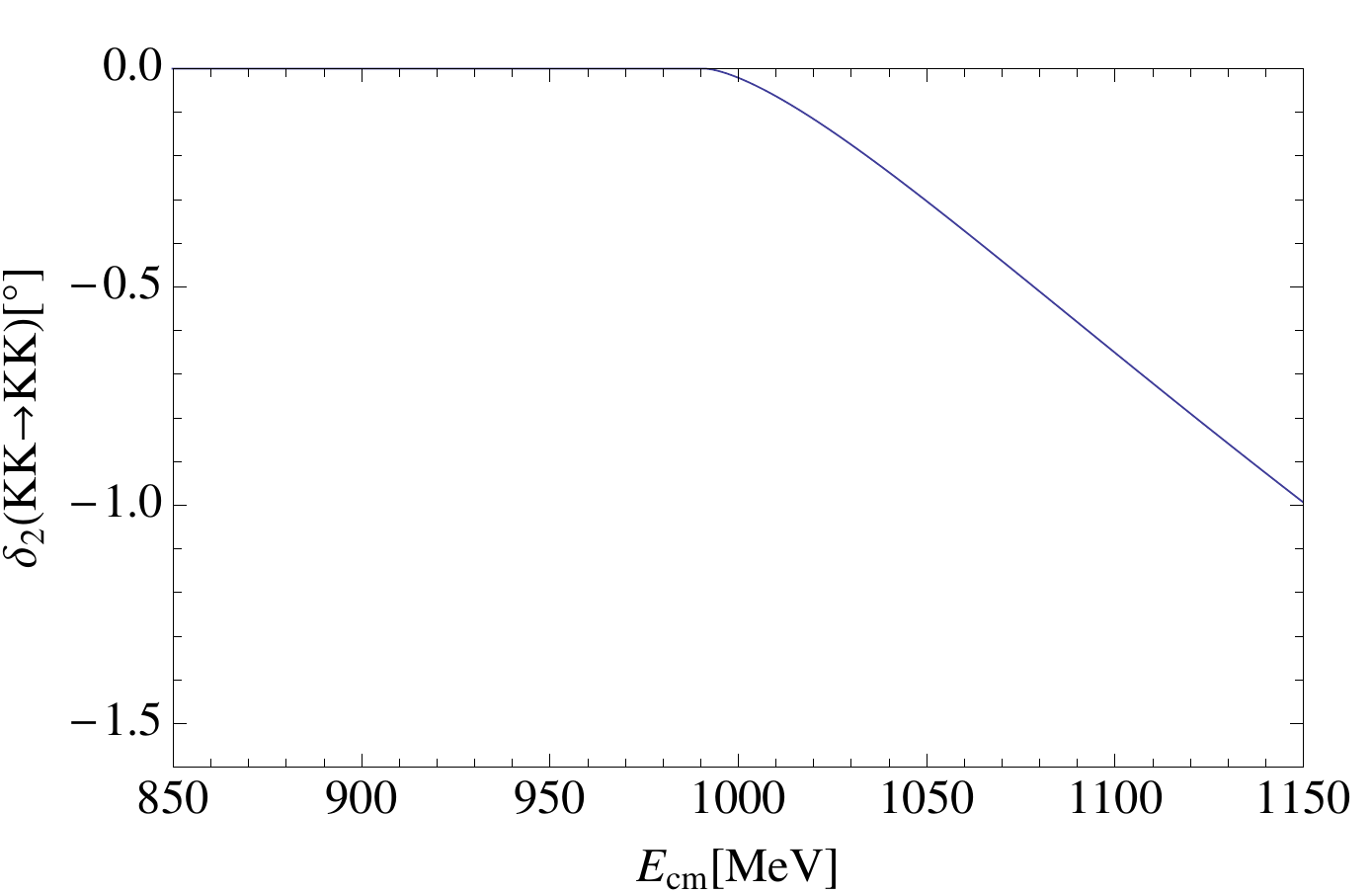}
\end{tabular}
\end{center}
\caption{Left: Elasticity of $\pi\pi\to\pi\pi$ at physical pion masses compared with experimental determinations~\cite{Protopopescu:1973sh}. 
The dashed line shows the inelasticity due to the $K\bar{K}$ channel alone as derived in reference~\cite{Niecknig:2012sj} 
from the Roy-Steiner solution in reference~\cite{Buettiker:2003pp}. 
Right: Phase shift $\delta_2(K\bar{K}\to K\bar{K})$. In this figure we only show the result of variant 2 discussed in the text (results for variant 1 are very similar).}
\label{fig:dkk}
\end{figure*}

The combined U$\chi$PT fit to eigenvalues at both pion masses is discussed in Sec.~\ref{sec:result}. 
We do not display the fit because it is almost indistinguishable from the blue curves in Fig.~\ref{fig:phase-shift}.
The result of the chiral extrapolation is shown in Fig.~\ref{fig:fitfincb} with the red band indicating the statistical uncertainties. The experimental data of Ref. \cite{Protopopescu:1973sh} are depicted with circles. As one can see, the $N_f=2$ extrapolation remains far from the experimental data. 

The two-channel U$\chi$PT formalism allows to estimate the effect of the missing strange quark in terms of the $K\bar{K}$ channel. For this, the $T$-matrix scattering amplitude, Eq. \ref{eq:tmat2}, is evaluated with the kernel $V$ from Eqs. \ref{eq:viam}, \ref{eq:v2c} and \ref{eq:v4c}. The LECs in the $\pi\pi\to\pi\pi$ transition $\hat{l}_1$ and $\hat{l}_2$ are fixed at their values from the combined fit to the $N_f=2$ lattice data (see Table \ref{table:uchiptfit1}).

The combinations of LECs appearing in the $\pi\pi\to K\bar K$ and the $K\bar{K}\to K\bar{K}$ transitions of Eq.~\ref{eq:v4c} are different from those
of $\pi\pi\to\pi\pi$ and taken from a global fit to $\pi\pi$ and $\pi K$
experimental phase shift data, similar as in Ref.~\cite{Doring:2013wka}. Statistical uncertainties from
this source are not considered, because they are smaller than those from lattice data. The relevant values from this fit are $L_3=-3.01(2)\times 10^{-3}$ and $L_5=0.64(3)\times 10^{-3}$, $\hat{l}_1=0.26(5)\times 10^{-3}$, $\hat{l}_2=-3.96(4)\times 10^{-3}$.

However, note that $\hat{l}_1$ and $\hat{l}_2$ also appear in the $\pi\pi\to K\bar K$ and the $K\bar{K}\to K\bar{K}$ transitions (see Eq. \ref{eq:v4c}).  
It is then not clear which values of $\hat l_i$ to use in these transitions. 
We have tested two variants:
\begin{enumerate}
\item
Evaluate the $\pi\pi\to K\bar{K}$ and $K\bar{K}\to K\bar{K}$ transitions with the $\hat{l}_1$ and $\hat{l}_2$ from the fit to $N_f=2$ lattice data.  
\item
Set all the LECs involved in the $\pi\pi\to K\bar{K}$ and $K\bar{K}\to K\bar{K}$ transitions to the LECs from the mentioned fit to experimental data.
\end{enumerate}
As Table~\ref{table:uchiptfit1} shows, the $\hat l_i$ from the fit to $N_f=2$ lattice data are similar to the ones quoted above, but not entirely compatible.

The result of the $2\to 3$ flavor extrapolation with these two variants is shown in Fig.~\ref{fig:fitfincb} with the two blue curves connected by the blue band.
The difference between these two strategies leads to about $20$ MeV difference in the mass of the $\rho(770)$ meson which gives an estimation of the uncertainties from model consistency.

Even though the $K\bar{K}$ channel has a 
significant impact on the mass of the $\rho$, the elasticity remains close to unity when 
this channel is open. This is shown in Fig.~\ref{fig:dkk} (left). The $K\bar{K}$-phase shift 
is small and negative, as shown in Fig.~\ref{fig:dkk} (right). It has the same sign as determined in Ref.~\cite{Wilson:2015dqa} 
at an unphysical pion mass.

Before we conclude, we want to address a possible concern regarding the four-pion channel, 
which we believe to have a negligible effect on our results and conclusions. 
The effect is expected to be small because the
branching ratio of $\rho$ to four pions is smaller than $2\times 10^{-3}$~\cite{Agashe:2014kda}.
The lattice simulation includes the four-pion channel and
the \uchipt\ model used here takes this effects into account only implicitly, through a shift
in the values of the fitted low energy constants. At the simulation     
points the effect is thus, at least approximately, included and the main uncertainty
comes from the chiral extrapolation. This is fundamentally different
from the $K$-$\bar K$ channel which is absent in the lattice simulation. We believe that the
uncertainty in the extrapolation is small, since the model
fits our data at two different pion masses consistently and the extrapolation
agrees very well, as can be seen from Fig.~\ref{fig:extrapolation}, with another $N_f=2$
lattice calculation very close to the physical point~\cite{Bali:2015gji}.

In summary, even with the discussed theoretical uncertainty, the shift of the $\rho$ mass by the $K\bar K$ channel is significant and leads to a surprisingly good post-diction of experiment.

\section{Extracted energies and correlation matrices}
\label{appendix:fitting}

In this section, we tabulate the details about fitting---fitting ansatz and fitting windows---for 
each energy level and for each ensemble used in this study. These details are reported in 
Table~\ref{Tab:fit_details}.
\begin{table*}[!htb]
\centering
\begin{tabular}{@{}*{7}{>{$}c<{$}}*{2}{>{\hspace{3mm}}l!{\hspace{3mm}}}*{1}{>{$0.}c<{$}}@{}}
\toprule
m_{\pi}(\MeV) & \mathbf{P} & \eta  &n & \text{ansatz}& t_0 & \text{fit window} & \multicolumn{1}{c}{aE} & \multicolumn{1}{c}{$\chi^2$/dof} & \multicolumn{1}{c}{Q}\\
\midrule
315 & (0,0,0) & 1.0   & 1 & d & 3 & 3-20 & 0.4932(16) & 0.65 & 83\\
    &         &       & 2 & d & 3 & 5-15 & 0.6612(14) & 0.61 & 77\\
    &         &       & 3 & d & 3 & 4-10 & 0.842(4) & 0.61 & 65\\
    &         & 1.25  & 1 & d & 3 & 4-15 & 0.4847(14) & 1.5 & 13\\
    &         &       & 2 & d & 3 & 5-14 & 0.5891(14) & 1.5 & 16\\
    &         &       & 3 & d & 3 & 4-12 & 0.785(5) & 0.82 & 55\\
    &         & 2.0   & 1 & dc & 3 & 4-16 & 0.4508(6) & 0.96 & 47\\
    &         &       & 2 & d & 3 & 6-17 & 0.5098(18) & 1.23 & 27\\
    &         &       & 3 & d & 3 & 5-15 & 0.6547(13) & 0.92 & 49\\
    &         &       & 4 & d & 3 & 3-12 & 0.704(2) & 0.09 & 99\\
\midrule
    & (0,0,1) & 1.0   & 1 & dt & 3 & 5-17 & 0.5024(8) & 0.64 & 77\\
    &         &       & 2 & d & 3 & 5-16 & 0.5768(15) & 0.47 & 89\\
    &         &       & 3 & d & 3 & 3-13 & 0.7492(15) & 0.30 & 96\\
    &         & 1.25  & 1 & dt & 3 & 7-15 & 0.4701(9) & 0.69 & 63\\
    &         &       & 2 & d & 3 & 5-16 & 0.547(2) & 0.95 & 47\\
    &         &       & 3 & d & 3 & 3-13 & 0.717(2) & 0.30 & 97\\
    &         & 2.0   & 1 & dt & 3 & 3-20 & 0.4241(7) & 0.95 & 50\\
    &         &       & 2 & d & 3 & 5-20 & 0.5036(9) & 1.36 & 17\\
    &         &       & 3 & d & 3 & 5-17 & 0.574(1) & 0.38 & 96\\
    &         &       & 4 & d & 3 & 3-13 & 0.676(1) & 0.37 & 94\\
\midrule  
226 & (0,0,0) & 1.0   & 1 & d & 3 & 3-15 & 0.4598(15) & 0.82 & 61\\
    &         &       & 2 & d & 3 & 3-12 & 0.6184(15) & 0.09 & 99\\
    &         &       & 3 & d & 3 & 3-8  & 0.820(8)   & 0.08 & 97\\
    &         & 1.17  & 1 & d & 3 & 3-12 & 0.448(2)   & 1.06 & 39\\
    &         &       & 2 & d & 3 & 3-12 & 0.558(2)   & 1.24 & 28\\
    &         &       & 3 & d & 3 & 3-10 & 0.744(18)  & 0.05 & 99\\
    &         & 1.33  & 1 & d & 3 & 3-13 & 0.4300(15) & 0.44 & 90\\
    &         &       & 2 & d & 3 & 3-14 & 0.527(2) & 0.35 & 96\\
    &         &       & 3 & d & 3 & 3-11 & 0.71(2)    & 0.08 & 99\\
\midrule
    & (0,0,1) & 1.0   & 1 & d  & 3 & 6-18 & 0.4217(9) & 0.68 & 75\\
    &         &       & 2 & d  & 3 & 5-16 & 0.5489(13) & 0.68 & 73\\
    &         &       & 3 & d  & 3 & 3-12 & 0.706(2)   & 0.19 & 99\\
    &         & 1.17  & 1 & d  & 3 & 6-20 & 0.391(1)   & 0.32 & 99\\
    &         &       & 2 & d  & 3 & 3-12 & 0.530(1)   & 0.88 & 52\\
    &         &       & 3 & d  & 3 & 3-10 & 0.672(3)   & 0.70 & 63\\
    &         & 1.33  & 1 & dt & 3 & 6-20 & 0.371(1) & 0.31 & 98\\
    &         &       & 2 & d  & 3 & 3-13 & 0.5095(11) & 0.97 & 46\\
    &         &       & 3 & d  & 3 & 3-11 & 0.656(2)    & 0.17 & 98\\
    &         &       & 4 & d  & 3 & 3-11 & 0.665(17)    & 1.09 & 26\\
\bottomrule
\end{tabular}
\caption{Extracted energies and fitting details. The ansatz labels the fitting function used; referring
to Section~\ref{sec:result} we have: {\it d} stands for double exponential function $f_1(t)$, 
{\it dc} is a double exponential function plus a constant term $f_2(t)$, and {\it dt} stands for
its boosted variant $f_3(t)$. $Q$ is the confidence level of the fit, that is the probability
under ideal conditions that the $\chi^2$ is larger than the fit result.}
\label{Tab:fit_details}
\end{table*}

As we discussed in Section~\ref{sec:fitting_method}, to determine resonance parameters
by fitting a functional description to our phase-shifts we need to take into account 
cross-correlations between the extracted energies. The energies extracted from different
ensembles are uncorrelated, but there will be correlations between the levels extracted
from the same ensemble. We computed these covariance matrices using a jackknife procedure.
These matrices are listed in Table~\ref{Tab:cov}.

\begin{table*}
\begin{tabular}{*{2}{>{$}l<{$}}}
\left(
\begin{array}{cccccc}
24.8     & 1.76     & 5.95   & 3.34    & 2.43  & 3.47\\
         & 20.4     & 14.7   & 4.04    & 6.74  & 12.4\\
         &          & 164    & 4.01    & 5.22  & 25.6\\
         &          &        & 6.8     & 2.67  & 4.16\\
         &          &        &         & 23.4  & 3.74\\
         &          &        &         &       & 24.4\\                 
\end{array}
\right) 
\times10^{-7}
&
\left(\begin{array}{cccccc}
24.0   &  -3.91   & -34.0   & 4.0   & 3.6   &-2.52\\
       & 23.8     & 71.0    & 3.57  &4.02   & 10.0\\
       &          & 760     &-3.77  & 23.5  & 59.0\\
       &          &         &8.71   & 1.18  & 3.17\\
       &          &         &       & 15.9  & -2.22\\
       &          &         &       &       & 41.0\\                 
\end{array}
\right)\times10^{-7} 
\\
\left(
\begin{array}{cccccc}
19.5     & -2.03    & -0.256  & 3.44   & 8.23   &-2.92 \\
         & 21.1     & 11.2    & 2.04   & 7.80   & 66.7 \\
         &          & 234     & 5.11   & 4.36   & 51.6 \\
         &          &         & 8.34   & 2.63   & 0.536\\
         &          &         &        & 30.0   & 0.269\\
         &          &         &        &        & 41.8\\                 
\end{array}
\right)\times10^{-7}
&
\left(
\begin{array}{cccccc}
33.9   & -14.7   & -44.4  & 1.09  & 5.67   & -10.8\\
       & 43.1    & 145    & 1.60  & 2.56   & 22.3\\
       &         & 2720   & 11.4  & 12.9   & 144 \\
       &         &        & 9.73  & 0.885  & 2.98\\
       &         &        &       & 12.1   & -0.436\\
       &         &        &       &        & 68.2\\                 
\end{array}
\right)\times10^{-7}
\\
\left(
\begin{array}{cccccccc}
3.62  & 2.24  & 1.54  & 1.87 & 3.16  & 3.10  & 1.90  & 1.74\\
      &32.4   & 0.195 & 3.15 & 1.71  & 8.42  & 2.34  & 0.884\\
      &       & 15.8  & 9.77 & 3.04  & -0.063& 9.99  & 6.21\\
      &       &       & 41.8 & 2.96  & 1.40  & 7.16  & 16.2\\
      &       &       &      & 4.29  & 1.48  & 2.97  & 2.20\\
      &       &       &      &       & 9.57  & 0.90  & 0.222\\
      &       &       &      &       &       & 10.1  & 4.62 \\
      &       &       &      &       &       &       & 9.78 \\                 
\end{array}
\right)\times10^{-7}
&
\left(
\begin{array}{ccccccc}
22.8  & -5.91  & -63.6  & 4.54  & 5.92  &-3.50  & -24.6\\
      & 42.8   & 108    & 3.80  &5.21   & 18.6  & 32.9 \\
      &        &5430    & 93.7  & -29.2 & 32.6  & 1690 \\
      &        &        & 19.3  & 2.78  & 1.62  & 57.4 \\
      &        &        &       & 14.2  & 2.45  & -31.6\\
      &        &        &       &       & 39.9  & -24.2\\                 
      &        &        &       &       &       & 2820 \\
\end{array}
\right)\times10^{-7}
\end{tabular}
\caption{Covariance matrices for the energies extracted from each ensembles. In the
left column from top down we list the ensembles ${\cal E}_{1,2,3}$ corresponding to $m_\pi\approx315\MeV$
and on the right ${\cal E}_{4,5,6}$ corresponding to $m_\pi\approx226\MeV$. The
order of the levels in each matrix corresponds to the order they appear in Table~\ref{Tab:fit_details}.} 
\label{Tab:cov}
\end{table*}

\newpage
\bibliography{ref}

\end{document}